\documentstyle[prd,aps]{revtex}
\tighten

\input epsf 

\begin{document}
\draft
\twocolumn[\hsize\textwidth\columnwidth\hsize\csname
@twocolumnfalse\endcsname
\title{AVOIDANCE OF COLLAPSE BY CIRCULAR \\
CURRENT-CARRYING COSMIC STRING LOOPS}
\author{Brandon CARTER$^1$, Patrick PETER$^1$ and Alejandro GANGUI$^{2,3}$}
\address{$^1$D\'epartement d'Astrophysique Relativiste et de Cosmologie,\\
Observatoire de Paris-Meudon, UPR 176, CNRS, 92195 Meudon (France),\\
$^2$ International Center for Theoretical Physics, \\
P.~O.~Box 586, 34100 Trieste, Italy, \\
$^3$ International School for Advanced Studies, \\ Via Beirut  4,
34014 Trieste,  Italy.\\
\sevrm Email: carter@obspm.fr, peter@prunelle.obspm.fr,
gangui@gandalf.sissa.it}
\maketitle
\begin{abstract}

Earlier attempts to calculate the nonlinear dynamical evolution of
Witten type superconducting vacuum vortex defects relied on the use of
approximate conducting string models that were too simple to take
proper account of the effect of current saturation. This effect is
however allowed for adequately in a newly developed class of rather
more complicated, though still conveniently analytic, conducting
string models. These more realistic models have recently been employed
by Larsen and Axenides for investigating the collapse of circular
string loops in the case for which angular momentum is absent. The
present work extends this investigation to the generic case of
circular string loops for which angular momentum is present, so that
there will be a centrifugal potential barrier. This barrier will
prevent collapse unless the initial conditions are such that the
relevant current saturation limit is attained, in which case the
string description of the vortex defect will break down, so that its
subsequent fate is hard to foresee. On the other hand if saturation is
avoided one would expect that the loop will eventually radiate away
its excess energy and settle down into a vorton type equilibrium
state.

\end{abstract}

\vskip2pc]

\section{Introduction}

Among the conceivable varieties of topological defects of the vacuum
that might have been generated at early phase transitions, the {\it
vortex type} defects describable on a macroscopic scale as {\it cosmic
strings} are the kind that is usually considered most likely to
exist. This is because even if they were formed at the Grand Unified
(GUT) scale, their density would be too low to induce a cosmological
catastrophe, contrary to what happens in the cases of domain walls and
monopoles~\cite{kibble,book}. However this consideration applies only
to the case of ordinary Goto-Nambu type strings, which ultimately
radiate away their energy and disappear. As was first pointed out by
Davis and Shellard~\cite{DS}, the situation is drastically different
for ``superconducting'' current-carrying strings of the kind
originally introduced by Witten~\cite{witten}. Indeed, it is becoming
clearer and clearer~\cite{vorton,96014} that the occurrence of stable
currents in strings can lead to a real problem because loops can then
be stabilized: the current, whether timelike or spacelike, breaks the
Lorentz invariance along the string
worldsheet~\cite{for,mal,neutral,enon0}, thereby leading to the
possibility of rotation~\cite{vorton}. The centrifugal effect of this
rotation may then compensate the tension in such a way as to produce
an equilibrium configuration, which, if it is stable, is what is known
as a {\it vorton}.  Whereas the energy density of non-conducting
string distribution decays like that of a radiation gas, in contrast a
distribution of relic vortons would scale like matter. Thus,
depending~\cite{96014} on when and how efficiently it was formed, such
a vorton distribution might eventually come to dominate the universe.

In view of this, it is very important to decide which rotating
equilibrium configurations really would be stable over cosmologically
significant timescales, and what fraction of the original population
of cosmic string loops would actually end up in such states. Dynamic
stability with respect to small perturbations has been
established\cite{stabgen,stab,stabwit} for most though not all of the
relevant equilibrium states within the framework of the classical
string description, but the question of stability against quantum
tunneling processes remains entirely open, being presumably dependent
on the postulated details of the underlying field theory. If the
currents were rigourously conserved the requirement that the
corresponding quantum numbers should lie in the range consistent with
stability would from the outset characterise the loops destined to
survive as vortons, but in practice things will be more complicated: a
lot of future work is needed to estimate the fraction of losses that
can be expected from mechanisms such as collisions, longitudinal
shocks, cusp formation, and occasional local violations of the
permissible current magnitude limits, that may occur before a
protovorton loop has finished radiating away its excess energy and
settled down as an actual vorton.

A loss mechanism of a rather extreme kind -- suggested originally for
non conducting strings by Hawking~\cite{SWH}, and considered more
recently in the context of conducting models by Larsen and
Axenides~\cite{LA} -- is that whereby a sufficiently large string loop
ends up by undergoing ``runaway collapse'' to form a black
hole. Events of this exotic kind are of intrinsic theoretical interest
in their own right, even though it is evident that they must be far
too rare to be of cosmological importance, since they can only occur
for very exceptional cases of initial dynamical configurations with an
extremely high degree of symmetry, meaning that they must be almost
exactly circular.

The investigation by Larsen and Axenides was restricted to the
reflection symmetric case characterised by absence of angular
momentum, for which they showed~\cite{LA}, subject to the neglect of
gravitational and electromagnetic self interaction, that the presence
of a current will not prevent an exactly circular loop from collapsing
to what in the framework of the string description would be just a
point, corresponding at a microscopic level to a configuration
compressed within the radius characterising the vacuum vortex core of
the string -- which will typically be of the order of the Compton
wavelength associated with the relevant Higgs mass, $m$ say -- so that
gravitational collapse would follow if the total mass-energy $M$ were
sufficiently large, $M\agt m^{-1}$ in Planck units. In such a case the
neglect\cite{LA} of an electromagnetic Coulomb barrier will
automatically be justifiable, not because it is entirely absent but
simply because it will be dominated by gravitational attraction,
provided the charge $Q$ (if any) on the loop is sufficiently small,
$Q\alt m^{-1}$. This condition will usually be satisfied because we
shall have $Q=Ze$ where $e$ is the relevant particle charge coupling
constant, which must either be zero if the current is of electrically
neutral type, or else must be equal to the electron charge $e\simeq
1/\sqrt{137} \approx 10^{-1}$, while $Z$ is the integral charge
quantum number: since the latter arises just from random fluctuations
it will seldom exceed the relevant limit $e^{-1} m^{-1}$, which is of
order $10^4$ in the GUT case $m\approx 10^{-3}$ and even higher for
lighter strings.

The present work extends the analysis of Larsen and Axenides~\cite{LA}
by treating generic circular states, for which the outcome is very
different.  Unlike the reflection symmetric zero angular momentum case
(which is the only possibility that can occur for a circular string
loop of the simple non-conducting kind) a generic circular state for a
conducting string loop will be subject to the centrifugal
effect. Whereas the Coulomb barrier will usually be negligible, on the
other hand the centrifugal barrier will usually be of dominant
importance. It is the centrifugal effect that makes possible the
existence of vorton type equilibrium states, and as will be seen below
the associated centrifugal barrier will generically prevent the kind
of collapse to a point that was envisaged by Larsen and Axenides. This
means that while such a collapse must be very rare even in the non
conducting string case previously envisaged by Hawking, it will be
much more extremely rare in the conducting string case envisaged here.

The motivation of the present work is not just to provide an explicit
quantitative demonstration of the qualitatively obvious phenomenon of
the existence of an infinite centrifugal barrier preventing the
collapse of a generic circular configuration of a conducting vortex
defect of the vacuum within the framework of the cosmic string
description. A less trivial purpose is to explore the limits of
validity of this {\it thin string} description by investigating the
conditions under which the current may build up to the saturation
limit beyond which the thin string approximation breaks down due to
local (transverse or longitudinal) instabilities -- so that a non
singular description of the subsequent evolution would require the use
of a more elaborate treatment beyond the scope of the present work. In
order to provide a physically complete analysis of such current
saturation phenomena, the present study needs to be generalised to
include non-circular configurations, whose treatment will presumably
require the use of numerical as opposed to analytic methods. This
consideration leads to a secondary motivation for the analytic
investigation provided here, which is to provide some firm results
that can be used for checking the reliability of the numerical
programs that are already being developed for the purpose of treating
conducting string loop dynamics in the general case.

There have been been many previous studies of circular conducting
string dynamics that -- unlike the recent analysis of Larsen and
Axenides~\cite{LA}, but like that of the present work -- already
included due allowance for the centrifugal effect. However these
earlier investigations were based on the use of conducting string
models that were too highly simplified to provide a realistic
description of Witten type vortex defects. The most obvious example of
such a highly simplified conducting string model is the linear type,
which has recently been applied to the case of circular loops by
Boisseau, Giacomini and Polarski~\cite{BGP}. A more elegant model
originally obtained from a Kaluza Klein type projection mechanism by
Nielsen~\cite{nielsen} has been used for studies of circular loops in
various contexts by several authors~\cite{several}, the application
that is most closely related to the present work being that of
Larsen~\cite{larsen}. The Nielsen model has the mathematically
convenient property of being transonic (meaning that transverse and
longitudinal perturbations travel at the same speed) and has been
shown to provide an accurate macroscopic description of the effect of
microscopic wiggles in an underlying Goto-Nambu type (non conducting)
string~\cite{CX}.  However this transonic model cannot describe the
physically important saturation effect that arises for large currents
in the more elaborate kind of string model~\cite{analytic} that is
needed for a realistic description of the essential physical
properties\cite{neutral} of a naturally occurring vacuum vortex such
as would result from the Witten mechanism~\cite{witten}.

\section{Current and the equation of state}

To describe a vacuum vortex defect by a cosmic string model, meaning
an approximation in terms of a structure confined to a two dimensional
worldsheet, it is necessary to know enough about the relevant
underlying field theoretical model to be able to obtain the
corresponding cylindrical (Nielsen Olesen type) vortex
configurations. The quantities such as the tension $T$ and the energy
per unit length $U$ that are needed for the macroscopic description in
terms of the appropriate thin string model are obtained from the
relevant underlying vortex configuration by integration over a
transverse section~\cite{neutral,enon0}. In simple non conducting
cases, the cosmic string models obtained in this way will be of the
Goto-Nambu type for which $T$ and $U$ are constant and equal to each
other. For more general vortex forming field theoretical models, the
corresponding cosmic string models will be characterised by variable
tension and energy which in a generic state will be related by an
inequality of the form $T<U$. In many such cases, and in particular in
the category envisaged by Witten~\cite{witten} -- to which the present
analysis like that of Larsen and Axenides~\cite{LA} is restricted --
the only independent internal structure on the string world sheet will
consist of a simple surface current $c^\mu$ say (which may or may not
be electrically charged), which implies that the dynamical behaviour
of the string model will be governed by an equation of state
specifying $T$ and $U$ as functions of the current magnitude
\begin{equation}
\chi=c^\mu c_\mu\ ,\label{chi}
\end{equation}
and hence as functions of each other~\cite{for,mal}.  In view of the
large number of different fields involved in realistic (GUT and
electroweak) field theoretical models it is not unlikely that an
accurate description of any vortex defects that may occur would
require allowance for several independent currents, but even if that
is the case one might expect that in typical situations one particular
current would dominate the others so that as a good approximation the
effects of the others could be neglected.

The following work, like that of Larsen and Axenides, is based on the
kind of string model~\cite{analytic} that is derivable on the basis of
Witten's pioneering approach\cite{witten} to the treatment of currents
in vacuum vortex defects. This approach is based on the plausible
supposition~\cite{neutral,enon0} that the essential large scale
features of such a phenomenon can be understood on the basis of an
appropriately simplified field theoretical model governed by an
effective action involving, in the simplest case just the gauged Higgs
type scalar field responsible for the local symmetry breaking on which
the very existence of strings depends, together with a complex scalar
``carrier'' field $\Sigma$, that is subject to a local or global
$U(1)$ phase invariance group, and that is confined to the vortex core
of the string with a phase that may vary along the string world sheet,
thereby determining a corresponding surface current. Such a Witten
type scalar field model is not only applicable to cases where the
underlying field responsible for the current is actually of this
simple scalar type: it can also provide a useful approximation for
fermionic fields~\cite{witten} as well as for vector
fields~\cite{alford,EW}. The carrier field will be expressible in the
form
\begin{equation}
\Sigma  = |\Sigma| \exp[i \varphi
\{\sigma ,\tau\}],
\end{equation}
with $\sigma$ and $\tau$ respectively the spacelike and timelike
parameters describing the string's worldsheet, where $\varphi$ is a
real phase variable, whose gradient will contain all the information
needed to characterise a particular cylindrical equilibrium
configuration of the vortex, and hence to characterise the local state
of the string in the cool limit for which short wavelength excitations
are neglected. In conceivable cases for which short wavelength
excitations contribute significantly to the energy a more elaborate
``warm'' string description would be needed\cite{warm}, but on the
basis of the assumption (which is commonly taken for granted in most
applications) that the cool limit description is adequate, it follows
that there will only be a single independent state parameter, $w$ say,
that can conveniently be taken~\cite{for,mal,analytic} to be
proportional to the squared magnitude of the gauge covariant
derivative of the phase with components $\varphi_{|a}$, using Latin
indices for the worldsheet coordinates $\sigma^{_1}=\sigma$,
$\sigma^{_2}=\tau$. We thus take the state parameter to be
\begin{equation}
 w =\kappa_{_0}\gamma^{ab}\varphi_{|a }\varphi_{|b} \ ,
\label{prim} \end{equation}
where $\kappa_{_0}$ is an adjustable positive dimensionless
normalisation constant, using the notation $\gamma^{ab}$ for the
inverse of the induced metric, $\gamma_{ab}$ on the worldsheet. The
latter will be given, in terms of the background spacetime metric
$g_{\mu\nu}$ with respect to 4-dimensional background coordinates
$x^\mu$, by
\begin{equation}
 \gamma_{ab}=g_{\mu\nu} x^\mu_{\, ,a} x^\nu_{\, ,b} \ ,
\end{equation}
using a comma to denote simple partial differentiation with respect to
the worldsheet coordinates $\sigma^a$. The gauge covariant derivative
$\varphi_{|a}$ would be expressible in the presence of a background
electromagnetic field with Maxwellian gauge covector $A_\mu$ by
$\varphi_{|a}=\varphi_{,a}\! -\! eA_\mu x^\mu_{\, ,a}$.  However in
the application developed below it will be assumed (as was done by
Larsen and Axenides~\cite{LA}) that the gauge term can be omitted,
either because the carrier field is uncoupled, meaning $e=0$, or else
because the electromagnetic background field is too weak to be
important which (as discussed in the introduction) will be a
sufficiently good approximation for most relevant applications, so
that it will be sufficient just to take $\varphi_{| a}$ to be the
simple partial derivative operation, $\varphi_{|a}=\varphi_{,a}$. With
even stronger justification it will also be assumed in the application
to be developed below that the local background gravitational field is
negligible, so that $g_{\mu\nu}$ can be taken to be flat.

Whether or not background electromagnetic and gravitational fields are
present, the dynamics of such a system will be governed~\cite{for,mal}
by a Lagrangian scalar, ${\cal L}$ say, that is a function only of the
state parameter $w$, and that determines the corresponding conserved
particle current vector, $z^a$ say, in the worldsheet, according to
the Noetherian prescription
\begin{equation}
z^a=- {\partial {\cal L}\over\partial \varphi_{|a} }\ ,
\end{equation}
which implies
\begin{equation}
{\cal K}z^a= \kappa_{_0} \varphi^{|a}\ ,
\label{zcur}\end{equation}
(using the induced metric for internal index raising) where ${\cal K}$
is given as a function of $w$ by setting
\begin{equation}
2{d{\cal L}\over dw}=-{1\over{\cal K}} \ .
\label{calk}\end{equation}
This current $z^a$ {\it in} the worldsheet can be represented by the
corresponding tangential current vector $z^\mu$ {\it on} the
worldsheet, where the latter is defined with respect to the background
coordinates, $x^\mu$, by
\begin{equation}
 z^\mu=z^a x^\mu_{\, ,a} \ .
\end{equation}

The purpose of introducing the dimensionless scale constant
$\kappa_{_0}$ is to simplify macroscopic dynamical calculations by
arranging for the variable coefficient ${\cal K}$ to tend to unity
when $w$ tends to zero, i.e. in the limit for which the current is
null. To obtain the desired simplification it is convenient not to
work directly with the fundamental current vector $z^\mu$ that (in
units such that the Dirac Planck constant $\hbar$ is set to unity)
will represent the quantized particle flux, but to work instead with a
corresponding rescaled particle current $c^\mu$ that is got by setting
\begin{equation}
z^\mu=\sqrt{\kappa_{_0}}\, c^\mu \ .
\label{scur}\end{equation}
In terms of the squared magnitude $\chi$ of this rescaled current
$c^\mu$, as given by (\ref{chi}) the primary state variable $w$ will
be given simply by
\begin{equation}
w={\cal K}^2\chi \ .
\end{equation}
It is to be remarked that in the gauge coupled case, i.e. if $e$ is
non zero, there will be a corresponding electromagnetic current vector
obtained by a prescription of the usual form $j^\mu=\partial {\cal
L}/\partial A_\mu$ which simply gives $j^\mu=e z^\mu$
$=e\sqrt{\kappa_{_0}} c^\mu$.

An important role is played in the theory by the dual Lagrangian,
$\Lambda$ that is obtainable~\cite{for} from the original Lagrangian
function ${\cal L}$ by a Legendre type transformation that gives
\begin{equation}
\Lambda={\cal L}+{\cal K}\chi\ .
\label{Lamb} \end{equation}
In the timelike current range where $w$ is negative the tension and
energy density will be respectively given by $T=-{\cal L}$,
$U=-\Lambda$, whereas in the spacelike current range where $w$ is
positive they will be given by $T=-\Lambda$, $U=-{\cal L}$. Local
stability requires the positivity of the squared speeds $c_{_{\rm
E}}^{\, 2}=T/U$ and $c_{_{\rm L}}^{\,2} =-dT/dU$ of extrinsic (wiggle)
and longitudinal (sound type) perturbations, so the admissible range
of variation of the state parameter $w$ -- or equivalently of the
squared current magnitude $\chi$ -- will be characterised by
\begin{equation}
{{\cal L} \over\Lambda}>0>{d{\cal L} \over d\Lambda}\ .
\label{stabc} \end{equation}

The appropriate function, ${\cal L}\{w\}$ for such a string model is
obtainable in principle by integrating the corresponding Lagrangian
scalar for the underlying field theoretical model over a two
dimensional section through the relevant cylindrical vortex
configuration. In practice this procedure can only be carried out with
high precision by using a numerical
treatment~\cite{neutral,enon0}. Progress was delayed for several years
by the difficulty of using the output of such a numerical treatment
for explicit dynamical applications. This problem has recently been
solved by the discovery of very simple empirical
formulae~\cite{analytic} (originally expressed using a systematic
notation scheme employing a tilde for duality, so that $\tilde\Lambda$
and $\tilde \chi$ represent what are respectively expressed here as
$\cal L$ and $-w$) that provide a convenient analytic description,
with sufficient accuracy for realism, within the limited range
(\ref{stabc}) of $w$ for which the string description is actually
valid.

The parameter $w$ can take both positive and negative values depending
on whether the current is spacelike or timelike, but for the Witten
vortex model that we consider here, it turns out that the
corresponding string description is valid only so long as it remains
within a bounded range~\cite{for,mal,neutral,enon0} -- outside which
vortex equilibrium states can still exist, but can no longer be
stable. What transpires~\cite{analytic} is that the effective
Lagrangian for the thin string description can be represented with
reasonably good accuracy throughout the allowed range (and with very
high accuracy in the timelike part for which $w<0$) by a function
${\cal L}$ that -- for a suitably adjusted (typically order of unity)
value of the normalisation constant $\kappa_{_0})$ -- is expressible
(even in the presence of electromagnetic and gravitational background
fields) in terms of just two independent parameters $m$ and $m_\ast$
in the form
\begin{equation}
{\cal L}\{w\}=-m^2 - {m_\ast^2\over 2} \ln
\left\{1+{w\over m_\ast^2 }\right\},
\label{Lag} \end{equation}
which leads to the very simple formula
\begin{equation}
{\cal K} =1+ {w\over m_\ast^2}
\label{kan} \end{equation}
for the function introduced above. The allowed parameter range
(\ref{stabc}) is specifiable by the condition that this function
should satisfy
\begin{equation}
\hbox{e}^{-2\alpha}<{\cal K}<2
\label{admissible}\end{equation}
where
\begin{equation}
\alpha=\big({m\over m_\ast}\big)^2\ ,
\label{massrat}\end{equation}
(The lower limit is where the tension $T$, and hence also the
extrinsic ``wiggle'' speed tends to zero, while the upper limit is
where the longitudinal perturbation speed tends to zero.) The fixed
parameters $m$ and $m_\ast$ have the dimensions of mass and can be
interpreted as expressing the respective orders of magnitude of the
relevant Higgs and the (presumably rather smaller) carrier mass
scales. It is to be noted that the work of Larsen and
Axenides~\cite{LA} was based on a previously proposed alternative
Lagrangian~\cite{analytic} that (in terms of the same parameters $m$
and $m_\ast$) provides a somewhat more accurate treatment of the
spacelike current range $w>0$, whereas the newer version~(\ref{Lag})
provides a treatment that is considerably more accurate for large
timelike currents.  For our present purpose the slight difference
between these alternative string models for Witten vortices is not of
qualitative physical importance: our main reason for preferring to use
the newer version (\ref{Lag}) has nothing to do with considerations of
very high precision, but is just that it turns out to provide more
conveniently explicit analytic expressions for the quantities that we
shall need.

\section{Conservation laws.}

The dynamical equations for such a string model are obtained from the
Lagrangian ${\cal L}$ in the usual way, by applying the variation
principle to a surface action integral of the form

\begin{equation}
 {\cal S} = \int d\sigma\,d\tau\,\sqrt{-\gamma}\, {\cal L}\{w\},
\label{action} \end{equation} (using the notation $\gamma\equiv
\det \{\gamma_{ab}\}$)
in which the independent variables are the phase field $\varphi$ on
the worldsheet and the position of the worldsheet itself, as specified
by the functions $x^\mu\{\sigma,\tau\}$.

Independently of the detailed form of the complete system, one knows
in advance, as a consequence of the local or global $U(1)$ phase
invariance group, that the corresponding Noether current will be
conserved, a condition which is expressible as
\begin{equation}
\big(\sqrt{-\gamma}\, z^a\big)_{,a}=0\ .
\end{equation}
For a closed string loop, this implies (by Green's theorem) the
conservation of the corresponding flux integral
\begin{equation}
Z=\oint d\sigma^a \epsilon_{ab} z^b\ ,
\label{zin}\end{equation}
where $\epsilon$ is the antisymmetric surface measure tensor (whose
square is the induced metric, $\epsilon_{ab}\epsilon^b{_ c}
=\gamma_{ac}$), meaning that for any circuit round the loop one will
obtain the same value for the quantum number $Z$, which is
interpretable as the integral value of the number of carrier particles
in the loop. The loop will also be characterised by a second
independent quantum number whose conservation is trivially obvious,
namely the topologically conserved phase winding number $N$ that is
defined by
\begin{equation}
2\pi N=\oint d\varphi=\oint d\sigma^a \varphi_{,a} \ .
\label{win}\end{equation}

As usual, the stress momentum energy density distribution $\hat
T{^{\mu\nu}}$ on the background spacetime is derivable from this
action by varying the background metric, according to the
specification
\begin{equation}
\hat T{^{\mu\nu}}\equiv {2\over\sqrt{-g}}{\delta{\cal S}\over
\delta g_{\mu\nu}}\label{tmunu} \equiv  {2\over\sqrt{-g}}
{\partial(\sqrt{-g}\,{\cal L})\over\partial g_{\mu\nu}} \ .
\end{equation}
This leads to an expression of the standard form
\begin{equation}
\sqrt{-g}\, \hat T^{\mu\nu}=\int d\sigma\,d\tau\,\sqrt{-\gamma}\,
\delta^{(4)} \left[x^\rho - x^\rho \{\sigma,\tau \}\right]\,
\overline T{^{\mu\nu}}. \end{equation}
in which the {\it surface} stress energy momentum tensor on the
worldsheet (from which the surface energy density $U$ and the string
tension $T$ are obtainable as the negatives of its eigenvalues) can be
seen to be given~\cite{for,mal} by
\begin{equation}
\overline T{^{\mu\nu}}={\cal L}\eta^{\mu\nu} +{\cal K} c^\mu c^\nu \ ,
\label{stress} \end{equation}
using the notation
\begin{equation}
\eta^{\mu\nu}= \gamma^{ab} x^\mu_{,a} x^\nu_{,b}
\end{equation}
for what is interpretable as the (first) fundamental tensor of the
worldsheet.

Independently of the particular form of the Lagrangian, the equations
of motion obtained from the action (\ref{action}) will be expressible
in the standard form~\cite{for,mal}
\begin{equation}
\overline\nabla_{\!\mu}\overline T{^\mu}{_\nu}=\overline f_\nu \ ,
\end{equation}
in which $\overline\nabla_{\!\mu}$ denotes the operator of surface
projected covariant differentiation, and where $\overline f_\mu$ is
the external force density acting on the worldsheet. When the effect
of electromagnetic coupling is significant this will be given in terms
of the field $F_{\mu\nu}=A_{\nu,\mu}-A_{\mu,\nu}$ by $\overline
f_\mu=eF_{\mu\nu} z^\nu$.  Even if this force density is non zero, its
contraction with the current vector $z^\mu$, or with the corresponding
rescaled current vector $c^\mu$, will vanish, and hence it can be seen
from the preceding formulae that the equations of motions
automatically imply the surface current conservation law
\begin{equation}
\overline\nabla_{\!\mu} c^\mu=0 \ ,
\end{equation}
which is the equivalent, in background tensorial notation, of the
condition expressed above in terms of $z^a$ using what was expressed
above in worldsheet coordinate notation. The background tensorial
operator $\overline\nabla$ in the foregoing equations is definable
formally by
\begin{equation}
\overline\nabla{^\mu}\equiv\eta^{\mu\nu} \nabla_{\!\nu}
\equiv x^\mu_{\, ,a}\gamma^{ab}\nabla_{\! b}
\end{equation}
where $\nabla$ is the usual operator of covariant differentiation with
respect to the Riemannian background connection. Thus for any closed
loop there will be a corresponding conserved circuit integral ${C}$
given by
\begin{equation}
{C}=\oint dx^\mu\varepsilon_{\mu\nu} c^\nu\ ,
\end{equation}
where $\varepsilon_{\mu\nu}$ is the background spacetime version of
the surface measure tensor $\epsilon_{ab}$, which means that its
contravariant version is the antisymmetric tangential tensor that is
given by
\begin{equation}
\varepsilon^{\mu\nu}=\epsilon^{ab} x^\mu_{\, ,a} x^\nu_{\, b}\ .
\end{equation}
This constant ${C}$ is of course just a rescaled version of the
integer particle quantum number $Z$, which will be given in terms of
it by
\begin{equation}
Z= \sqrt{\kappa_{_0}}\, {C} \ .
\label{zed} \end{equation}

In the following work (as in the preceding work of Larsen and
Axenides~\cite{LA}) it will be assumed either that the current is
uncoupled or else (as will more commonly be the case) that
$F_{\mu\nu}$ is negligible, so that we can simply take
\begin{equation}
\overline f_\mu=0\ .
\end{equation}

As well as neglecting electromagnetic correction effects, we shall now
also restrict our attention to cases in which the background is both
axisymmetric and stationary, as is the case for the flat space in
which we are in the end most particularly interested.  This means that
there will be corresponding vectors, ${\ell}^\mu$ and $k^\mu$ say,
that satisfy the Killing equations
\begin{equation}
\nabla_{\!\mu}{\ell}_\nu+\nabla_{\!\nu}{\ell}_\mu=0\ ,\hskip 1 cm
\nabla_{\!\mu}k_\nu+\nabla_{\!\nu}k_\mu=0\ ,
\label{killing} \end{equation}
and that will be respectively interpretable as generators of rotations
and time translations, so that when suitably normalised, their effect
can be expressed in the form
\begin{equation}
k^\mu{\partial\over\partial x^\mu}={\partial \over\partial t}\ ,\hskip
1 cm {\ell}^\mu{\partial\over\partial
x^\mu}=2\pi{\partial\over\partial \phi}\ ,
\label{killnorm} \end{equation}
where $t$ is an ignorable time coordinate and $\phi$ is an ignorable
angle coordinate. This normalisation is such that the total
circumferential length of the circular trajectory of the angle Killing
vector will simply be given by ${\ell}$ where
\begin{equation}
{\ell}^2={\ell}^\mu {\ell}_\mu \ .
\label{circ}\end{equation}

These Killing vectors can be employed in the usual way to define the
corresponding angular momentum surface current vector ${\cal J }^\mu$,
and the corresponding energy current vector ${\cal E}^\mu$, by setting
\begin{equation}
\overline T{^\mu}{_\nu}{\ell}^\nu\ =2\pi {\cal J}^\mu,\hskip 1 cm
\overline T{^\mu}{_\nu}k^\nu\ =-{\cal E}^\mu\ ,
\label{fluxes} \end{equation}
These currents will then satisfy surface conservation laws
\begin{equation}
\overline\nabla_{\!\mu}{\cal J}^\mu=0\ , \hskip 1 cm
\overline\nabla_{\!\mu}{\cal E}^\mu=0\ ,
\end{equation}
that have the same form as that satisfied by the current $c^\mu$.
This means that for a closed loop there will be corresponding
conserved angular momentum and mass-energy integrals, $J$ and $M$ say,
that will be given by
\begin{equation}
 J=\oint dx^\mu\varepsilon_{\mu\nu} {\cal J}^\nu \ ,\hskip 1 cm
M=\oint dx^\mu\varepsilon_{\mu\nu} {\cal E}^\nu\ .
\label{constants} \end{equation}

\section {Constants of circular motion}

We now restrict ourselves to cases for which the string configuration
itself shares the background spacetime property of being symmetric
with the action generated by the Killing vector ${\ell}^\mu$. This
entails that ${\ell}^\mu$ should be tangential to the worldsheet, i.e.
\begin{equation}
{\ell}^\mu=\lambda^a x^\mu_{\, ,a}
\end{equation}
where $\lambda^a$ is a corresponding Killing vector with respect to
the intrinsic geometry of the worldsheet, which -- on the
understanding that ${\ell}^\mu$ is interpretable, in the manner
described above, as the generator of angular rotation about a symmetry
axis -- means that the string configuration is {\it circular}, its
circumference at any instant being given by the local value of
${\ell}$.

In such a case, this Killing vector ${\ell}^\mu$ can be used to
generate the -- in that case circular -- circuit used for evaluating
these integrals, i.e. the infinitesimal displacement in the integrand
can be taken to be given by $2\pi\, d\sigma^a=\lambda^a d\phi$ so that
we obtain
\begin{equation}
2\pi\, dx^\mu= x^\mu_{\, ,a} \lambda^a d\phi = {\ell}^\mu d\phi ,
\label{circuit} \end{equation}
where $\phi$ is an ignorable angle coordinate of the usual kind with
period $2\pi$ as introduced above. (Such an angle coordinate can be
conveniently used to specify the first worldsheet coordinate,
$\sigma^{_1}$, by setting $\sigma=\phi$, so that by taking the second
world sheet coordinate, $\sigma^{_2}=\tau$ to be constant on the
circular symmetry trajectories, the components of the intrinsic
Killing vector are obtained in the form $\{\lambda^{_1},
\lambda^{_2}\}=\{1,0\}$.) Substituting this ansatz for $dx^\mu$ in the
corresponding integral formulae, it can be seen that the global integrals
(\ref{zin}) and (\ref{win}) for the winding number $N$ and the
particle number $Z$ will be given directly in the circular case by
corresponding {\it locally defined} Bernoulli type constants of the
motion ${B}$ and ${C}$ say, according to the relations
\begin{equation}
{B}= 2\pi\sqrt{\kappa_{_0}}\, N\ ,\hskip 1 cm
{C}= {Z\over\sqrt{\kappa_{_0}}}\ .
\label{proportion}\end{equation}
where these quantities -- of which the latter, ${C}$ is directly
identifiable with the global flux of the current $c^\mu$ so that it is
justifiable to designate it by the same symbol -- are now to be
thought of as being constructed according to the prescriptions of the
purely local form~\cite{CFH}
\begin{equation}
{B}=\sqrt{\kappa_{_0}}\,\lambda^a \varphi_{|a}\ ,\hskip 1 cm
{C}={\ell}^\mu\varepsilon_{\mu\nu}c^\nu\ .
\label{Bernoulli} \end{equation}
The reason why the single symmetry generator ${\ell}^\mu$ gives rise
to not just one but two independent Bernoulli type constants in this
way is attributable to the the string duality property~\cite{for,mal}.
(It is to be noted that instead of using the rationalised constants
${B}$ and ${C}$, the work of Larsen~\cite{LA,larsen} uses
corresponding unrationalised constants $n$ and $\Omega$ that are
expressible in terms of our present notation by $\Omega=-{C}/2\pi$ and
$n=\sqrt{\kappa_{_0}}N={B}/2\pi$.)

In a similar manner the mass (when it is defined) and angular momentum
integrals introduced in the previous section will also be expressible
in terms of purely locally defined constants of the motion in the
circular case. To do this it is convenient~\cite{CFH} to start by
introducing an effective momentum tangent vector ${\mit \Pi}^\mu$
given in terms of the relevant Killing vector, namely ${\ell}^\mu$ in
the case with which we are concerned, by the ansatz
\begin{equation}
{\mit \Pi}^\mu={\ell}^\nu\varepsilon_{\nu\rho}\overline T{^{\rho\mu}}\
.
\label{ansatz} \end{equation}
Since the surface stress momentum energy tensor, $\overline
T{^{\mu\nu}}$, only has components in directions tangential to the
worldsheet, it is evident that this formula provides a vector ${\mit
\Pi}^\mu$ that automatically has the required property of being
tangent to the string worldsheet.

Whenever the background spacetime is invariant under the action of
another Killing vector $k^\mu$ -- which in the application to be
considered here will be interpreted as expressing stationarity -- so
that the corresponding integral formula (\ref{constants}) for $M$ is
well defined, it can be seen using (\ref{circuit}) again that this
globally defined quantity will now be obtainable as a purely local
constant of the motion from the formula
\begin{equation}
 k^\mu{\mit \Pi}_\mu=-M \ .
\label{encon} \end{equation}
It can also be seen that whether or not the background has a
(stationary) symmetry generated by $k^\mu$, the other (angular
momentum) constant $J$ provided according to (\ref{constants}) by the
original (axisymmetry) Killing vector ${\ell}^\mu$ itself will
similarly be obtainable, in the circular case, as a purely local
constant of the motion given by
\begin{equation}
{\ell}^\mu{\mit \Pi}_\mu = 2 \pi J\ .
\label{angcon} \end{equation}
It is however to be noted that this last constant is not independent
of the ones presented above: it can be seen by substituting the
formula (\ref{stress}) in (\ref{ansatz}) and using the defining
relations (\ref{zcur}) and (\ref{scur}) for the current that it will
be expressible in terms of the two (mutually dual) Bernoulli constants
(\ref{Bernoulli}) by the simple product formula
\begin{equation}
J={{B}{C}\over 2\pi} =NZ \ .
\label{angquant} \end{equation}
This shows that the integral quantisation of the winding and particle
numbers $N$ and $Z$ automatically entails the integral quantisation of
the angular momentum $J$.

\section{Dynamics of circular motion}

Whenever the string motion shares the symmetry generated by a
background spacetime Killing vector, the problem of solving the
equations of motion for its two dimensional worldsheet can naturally
be reduced to a problem of finding a one dimensional trajectory
tangential to the worldsheet but not aligned with the symmetry
generator, since when such a trajectory has been found it is trivial
to extend it to the complete two dimensional world sheet by the
symmetry action. The general procedure for obtaining such a tangential
trajectory for symmetric solutions of the equations of motion of
conducting string models was originally developed by Carter, Frolov,
and Heinrich~\cite{CFH}, who applied this method to study stationary
solutions in a Kerr black hole background. This method was adapted to
the kind of situation with which the present application is concerned,
namely circular instead of stationary symmetry, by
Larsen~\cite{larsen}. The original derivation~\cite{CFH} involved the
use of the quotient space with respect to the relevant symmetry
action, but a more recent and general treatment~\cite{mal} (including
allowance for the possibility of axionic as well as electromagnetic
coupling) has provided a more direct route that does not need such an
auxiliary construction.

What this procedure provides is a particular kind of world sheet
generating trajectory that is characterised by having a tangent vector
${\mit \Pi}^\mu$ given in terms of the relevant Killing vector, namely
${\ell}^\mu$ in the case with which we are concerned, by the ansatz
(\ref{ansatz}). The procedure makes use of the fact~\cite{mal,CFH}
that, for given values of the Bernoulli constants ${B}$ and ${C}$, the
state function $w$, and hence also the squared magnitude of the
tangent vector ${\mit \Pi}^\mu$, namely the quantity
\begin{equation}
{\mit \Psi}={\mit \Pi}^\mu {\mit \Pi}_\mu \ ,
\end{equation}
(which will play the role of a potential) can be specified in advance
as a {\it scalar field} over the entire background space (not just a
single string worldsheet), in such a way as to agree with the
respective physical values of ${\mit \Psi}$ on the particular
worldsheet under consideration.

In the generic case this is done by expressing ${\mit \Psi}$ as a
function of one of the independent state variables, $w$ say, which
will itself be expressible, as a function of the squared Killing
vector magnitude ${\ell}^2$, and hence as a scalar field over the
entire background space by solving the equation
\begin{equation}
{\ell}^2={{B}^2\over w}-{{C}^2\over\chi} \ ,
\label{state} \end{equation}
where $\chi$ is the squared magnitude (\ref{chi}) of the current
vector, which is obtainable from the Lagrangian $\cal L$ as a function
of $w$ in the manner described above. It is to be remarked that in the
recent derivation\cite{mal} the corresponding formula (expressed using
the notation $\beta$ for ${C}$, $\tilde\beta$ for ${B}$, and
$-\tilde\chi$ for $w$) contains a transcription error that has the
effect of replacing $w$ and $\chi$ by their respective squares.
Except in the ``chiral'' -- i.e. {\it null current} -- limit for which
$w$ and $\chi$ vanish so that a special treatment is needed, the self
dual formula (\ref{state}) is immediately obtainable by using the
Bernoulli formulae (\ref{Bernoulli}) to evaluate the respective
components of ${\ell}^\mu$ parallel to and orthogonal to the current.

Having used this procedure to obtain $w$ as a field over the
background spacetime, one can then use the result to obtain the
corresponding value of the squared magnitude ${\mit \Psi}$ -- which,
in the recent derivation~\cite{mal} was written as $X^2$, though it
need not be positive, since the vector ${\mit \Pi}^\mu$ may be
timelike or null as well as spacelike).  This quantity will be
obtainable using (\ref{state}) as a function of the squared Killing
vector magnitude ${\ell}^2$ (which in the case of symmetry must
necessarily be positive) by the manifestly self dual formula
\begin{equation}
{\mit \Psi}={{C}^2\Lambda^2\over\chi}-{{B}^2{\cal L}^2\over w}\ ,
\label{psdual}\end{equation}
where $\Lambda$ is the dual Lagrangian function, as given by
(\ref{Lamb}).

In order to obtain a treatment that remains valid in the ``chiral''
limit -- for which $w$ and $\chi$ vanish, so that the formulae
(\ref{state}) and (\ref{psdual}) become indeterminate -- it is
convenient to rewrite ${\mit \Psi}$ in a manner that sacrifices
manifest self duality by expressing it in the form
\begin{equation}
{\mit \Psi}={{B}^2{C}^2\over{\ell}^2}-{\mit \Upsilon}^2\ ,
\end{equation}
where (unlike the quantity $X=\sqrt{{\mit \Psi}}$ which may be
imaginary) the quantity ${\mit \Upsilon}$ defined in this way will
always be real in the admissible range (\ref{stabc}), as can be seen
by expressing it in either of the equivalent -- mutually dual --
alternative forms
\begin{equation}
{\mit \Upsilon}= {{B}^2\over{\cal K}{\ell}}-\Lambda{\ell}=
{{C}^2 {\cal K}\over{\ell}}-{\cal L}{\ell}\ ,
\label{enfunc} \end{equation}
of which the latter is the most convenient for practical calculations
starting from a given form of the Lagrangian ${\cal L}$.

In the generic case, $B^2\neq C^2$, the required quantities ${\cal K}$
and ${\cal L}$ are obtained indirectly as functions of ${\ell}$ by
solving (\ref{state}) which will give a result that is always
non-null, i.e $w$ and $\chi$ will never pass through zero, so the
current will preserve a character that is permanently timelike or
permanently spacelike as the case may be. The exception is the
``chiral'' case, which is characterised by the equality $B^2=C^2$, and
for which the only possible states are of the null kind characterised
by $w=\chi=0$, so that the required quantities ${\cal K}$ and ${\cal
L}$ will be given directly, independently of ${\ell}$, as the
constants ${\cal K}=1$ and ${\cal L}=-m^2$.

It is of particular interest for the dynamical applications that
follow to obtain the derivative of the field ${\mit \Upsilon}$ with
respect to the cylindrical radial coordinate ${\ell}$ with respect to
which it is implicitly or (in the chiral case) explicitly defined: it
is obvious in the chiral case since in this case $\Lambda = {\cal L}$,
and can also be verified (using the relation
\begin{equation}
{d{\ell}\over dw} = -{{\ell}\over 2 w} + {C^2 (d{\cal K}/dw)\over
2{\ell} w
(d{\cal L}/dw)}
\end{equation}
for the variations of ${\ell}$) in the generic case for which ${\cal
L}$ and ${\cal K}$ are variable, that this derivative will be
expressible in either of the very simple -- mutually dual --
equivalent forms
\begin{equation}
{d{\mit \Upsilon}\over d{\ell}}=-{C^2 {\cal K}\over{\ell}^2}-\Lambda
=-{B^2\over {\cal K}{\ell}^2}- {\cal  L}
\label{deri} \end{equation}
of which again it is the latter that is most convenient for practical
calculations starting from a given form of the Lagrangian ${\cal L}$.

After having obtained the field ${\mit \Psi}$ in this way, the final
step in the procedure for obtaining the string tangent vector ${\mit
\Pi}^\mu$ is to integrate its equations of motion, which can easily be
shown~\cite{mal,CFH,brane} to have the form
\begin{equation}
2{\mit \Pi}^\nu\nabla_{\!\nu}{\mit \Pi}_\mu= \nabla_\mu {\mit \Psi} \ ,
\label{congeod} \end{equation}
subject to the constraint
\begin{equation}
{\ell}^\mu {\mit \Pi}_\mu={B}{C}\ ,
\label{momcon} \end{equation}
whose conservation is an automatic consequence of the symmetry. This
constant is interpretable according to (\ref{angquant}) as being
proportional to the angular momentum $J$.

The foregoing formulation depends only on the existence of the
symmetry generated by ${\ell}^\mu$, which is postulated to apply not
only to the background but also to the string solution itself -- so
that the further supposition that this symmetry is that of rotation
about an axis means that the string configuration is circular.  It
does not depend on the additional postulate that the background space
time is also subject to another symmetry generated by the independent
Killing vector, $k^\mu$ that is responsible for the existence of the
mass-energy constant $M$ given by (\ref{encon}).

In order to solve the equation of motion (\ref{congeod}), which can be
seen to be interpretable as that of a geodesic with respect to a
conformal modified background metric given by ${\mit \Psi}
g_{\mu\nu}$, it is useful to employ a Hamiltonian formulation of the
standard form
\begin{equation}
{dx^\mu\over d\tau}={\partial H\over\partial {\mit \Pi}_\mu}\ ,
\hskip 1 cm {d{\mit \Pi}_\mu\over d\tau}= -{\partial H\over
\partial x^\mu}\ ,
\label{Hameq} \end{equation}
where $\tau$ is parameter along the trajectory, that can conveniently
be used to specify the choice of the second internal coordinate
$\sigma^{_2}$ (the first one, $\sigma^{_1}$, having already been
chosen to be the angle coordinate $\phi$). Such a formulation of the
conformal geodesic equation is readily obtainable by taking the
Hamiltonian to be given by
\begin{equation}
2H=g^{\mu\nu}{\mit \Pi}_\mu{\mit \Pi}_\nu-{\mit \Psi} \ ,
\label{Hamilt} \end{equation}
with the understanding that the system is to be solved subject to the
constraint
\begin{equation}
H=0\ ,
\label{Hamcon}\end{equation}
which ensures the correct normalisation of the tangent vector, which
by the first Hamiltonian equation will be given directly by
\begin{equation}
{\mit \Pi}^\mu={dx^\mu\over d\tau} \ .
\end{equation}

\section{Solution for the state function.}

In order to carry out the procedure summarised in the preceding
section, we first have to solve the equation (\ref{state}) for the
state variable $w$. In terms of the magnitude ${\ell}$ of the
axisymmetry generator ${\ell}^\mu$, which will give the local value of
the circumference of the circular string loop, this equation will be
expressible as
\begin{equation}
{\ell}^2 w={B}^2-{C}^2{\cal K}^2\ .
\end{equation}

For a general conducting string model this equation would be hard to
solve explicitly, but there are special cases for which a convenient
analytic solution is available, the first known example being that of
the transonic string model, for which the equation for the field $w$
is found to be simply {\it linear}, so that it can immediately be
solved to provide a system that turns out to be completely integrable
by separation of variables in a Kerr black hole background when the
symmetry under consideration is stationarity~\cite{CFH}, though
unfortunately not when it is axisymmetry~\cite{larsen}, in which case
such complete integrability is available only for purely equatorial
configurations, including in the limit, the kind of flat space ring
configurations with which the present study is ultimately
concerned. Another case whose application to circular configurations
has recently been considered~\cite{BGP}, and in which the equation
(\ref{state}) is also simply linear, is that of the even cruder fixed
trace model (for which ${\cal K}$ is just a constant) that was
originally suggested by Witten himself~\cite{witten} to describe the
effect of his mechanism for currents that are very small compared with
saturation, but which turns out~\cite{neutral} to be misleading
(because subsonic) even in that limit.

The more recent work of Larsen and Axenides~\cite{LA} was more
advanced in that it used the newer kind of string model\cite{analytic}
that (unlike the simple transonic model and its cruder fixed trace
predecessor) can provide a realistic account of the current saturation
effect that is a salient feature of the Witten mechanism; this work
was however much more specialised than the preceding investigations
cited above, as it only considered the non rotating case of vanishing
winding number $N$, which in our present notation means ${B}=0$. As
well as allowing for non-zero winding number, the present work
involves a physically unimportant but technically valuable improvement
in that we use more recently proposed\cite{analytic} string model
characterised by (\ref{Lag}), which turns out to be particularly
convenient for the present stage in our analysis since it leads to an
equation for $w$ that although not actually linear, as was the case
for the transonic model, can be seen to be the next best thing,
meaning that it is just {\it quadratic} (whereas the version used by
Larsen and Axenides gives an equation for $w$ that has a much more
awkward quartic form).  The result of using (\ref{Lag}) is expressible
in the form
\begin{equation}
{C}^2{\cal K}^2+m_\ast^{\, 2}{\ell}^2\big({\cal K}-1\big)-{B}^2=0\ ,
\label{quadratic} \end{equation}
which can immediately be solved to give
\begin{equation}
{\cal K}={-m_\ast^{\, 2}{\ell}^2
+\sqrt{4{C}^2({B}^2+m_\ast^{\,2}{\ell}^2)+m_\ast^{\, 4}{\ell}^4}
\over 2{C}^2} \ ,
\label{ksolution} \end{equation}
choosing the positive sign for the square root because ${\cal K}$ is
positive throughout the admissible range (\ref{stabc}) for the state
parameter $w$.

In terms of this explicit formula for ${\cal K}$ the state function
$w$ itself is immediately obtainable using the expression
\begin{equation}
w= m_\ast^{\, 2}\big({\cal K} -1\big) \
\end{equation}
that applies for this model.  Since our Lagrangian (\ref{Lag}) can be
expressed directly in terms of ${\cal K}$ as
\begin{equation}
{\cal L}=-m_\ast^{\, 2}\big(\alpha+ \ln\sqrt{{\cal K}}\big)\ ,
\end{equation}
our explicit formula for ${\cal K}$ can also be directly applied to
obtain the required potential ${\mit \Psi}$, for which we obtain the
formula
\begin{equation}
{\mit \Psi}={{B}^2{C}^2\over {\ell}^2}-{\ell}^2\Big({{C}^2{\cal
K}\over{\ell}^2} +m_\ast^2(\alpha+\ln\sqrt{\cal K})\Big)^2\ .
\end{equation}

\section{Motion in a flat background}

Up to this point we have been using a formulation that is valid for an
arbitrary stationary axisymmetric background, including for example
that of a Kerr black hole. In order to obtain a result that is
completely integral in explicit form, and because it is the case of
greatest physical importance, we shall now restrict our attention to
the case of a flat space background, for which there will be no loss
of generality in supposing the circular string loop to be confined to
an equatorial hyperplane with 3-dimensional spacetime metric given in
terms of circular coordinates $\{r,\phi,t\}$ by
\begin{equation}
ds^2= dr^2+r^2 d\phi^2-dt^2 \ ,
\label{flat}\end{equation}
so that the Killing vectors used in the discussion above will be
identifiable as $\{k^{_1},k^{_2},k^{_3}\} =\{0,0,1\}$ and
$\{{\ell}^{_1},{\ell}^{_2},{\ell}^{_3}\} =\{0,2\pi,0\}$.

In these circumstances the circumferential length field ${\ell}$ that
played a fundamental role in the preceding discussion will be given
simply by
\begin{equation}
{\ell}=2\pi r \ ,
\end{equation}
and the evolution of the circular string worldsheet will be given
simply by specifying the radius $r$ as a function of the background
time $t$. We shall use a dot to denote differentiation with respect to
this time $t$, which will vary proportionally to the Hamiltonian time,
$\tau$, with coefficient given by the energy constant, so that we
shall have
\begin{equation}
 {dt\over d\tau}=M  \ .
\label{tdot}\end{equation}

For a complete physical description of the solution, it would also be
necessary to specify the distribution over the worldsheet of the phase
field $\varphi$, which must evidently have the form
\begin{equation}
\varphi=q+N\phi \ ,
\end{equation}
where $N$ is the conserved winding number as defined above and $q$ is
a function only of $t$.

The fact that the fourth (azimuthal) direction can be ignored in this
particularly simple case means that the complete set of equations of
motion is provided directly in first integrated form by the constants
of the motion.  Using the formulae (\ref{zcur}) and (\ref{scur}) to
work out the expression (\ref{Bernoulli}) for the Bernoulli constant
${C}$, it can be seen that the time derivative of this function $q$
will be given in terms of that of the radius $r$ by
\begin{equation}
\dot q={C}{\cal K}{\sqrt{1-\dot r^2}\over 2\pi r\sqrt{\kappa_{_0}}}\ .
 \end{equation} By similarly using the formula (\ref{stress}) to work
 out the expression (\ref{encon}) for the mass-energy constant $M$ the
 evolution equation for $r$ can be obtained in the first integrated
 form
\begin{equation}
M\sqrt{1-\dot r^2}={\mit \Upsilon}
\label{radevol}\end{equation}
where ${\mit \Upsilon}$ is the quantity that is given by the formula
(\ref{enfunc}), whose evaluation as a function of the circumference,
${\ell}=2\pi r$ is discussed in Section V.

Instead of going through the detailed evaluation of the expression
(\ref{encon}) using (\ref{stress}), a more elegant albeit less direct
way of obtaining the same equation of motion for $r$ is to apply the
Hamiltonian formalism described in Section V. It is evident from
(\ref{tdot}) that in the flat background (\ref{flat}) the radial
momentum component ${\mit \Pi}_{_1}$ will be given by
\begin{equation}
{\mit \Pi}_{_1}={dr\over d\tau}=M \dot r \ ,
\end{equation}
and under these conditions the Hamiltonian (\ref{Hamilt}) will reduce
to the simple form
\begin{equation}
 H={1\over 2}\left({\mit \Pi}_{_1}^{\, 2}+{J^2\over r^2}-M^2-{\mit
 \Psi}\right)\ .
\end{equation}
It is to be remarked that the term $J^2/r^2$ in this formula has the
form of the centrifugal barrier potential that is familiar in the
context of the analogous problem for a point particle. By what is a
rather remarkable cancellation, it can be seen that the effect of the
extra potential ${\mit \Psi}$ taking account of the elastic internal
structure of the string is merely to replace the familiar centrifugal
barrier contribution $J^2/r^2$ by a modified barrier contribution
given simply by ${\mit \Upsilon}^2$ where ${\mit \Upsilon}$ is the
scalar field (\ref{enfunc}) introduced in the previous section, since
it can be seen that the relevant combination of terms turns out to be
expressible simply as
\begin{equation}
 {J^2\over r^2}-{\mit \Psi}={\mit \Upsilon}^2
\end{equation}
The normalisation expressed by the constraint that the Hamiltonian
should vanish can thus be seen to give the equation of motion for $r$
in the convenient first integrated form
\begin{equation}
M^2 \dot r^2=M^2 -{\mit \Upsilon}^2 \ , \end{equation} which is
evidently equivalent to the radial evolution equation (\ref{radevol})
given above.

\section {Stationary ``vorton'' states}

As an immediate particular consequence of this equation of motion, it
can be seen that there will be {\it vorton type} equilibrium
solutions, with mass energy given by
\begin{equation}
M={\mit \Upsilon} \ ,
\end{equation}
wherever the relevant effective energy function ${\mit \Upsilon}$
satisfies the stationarity condition
\begin{equation}
{d{\mit \Upsilon}\over dr}=0 \ .
\label{equilibrium} \end{equation}

The formula (\ref{deri}) given above for the derivative of ${\mit
\Upsilon}$ can be used to write this stationarity condition in the
form
\begin{equation}
C^2{\cal L}w=B^2\Lambda\chi \ .
\end{equation}
This is recogniseable as the equilibrium requirement that is well
known from previous more specialised studies of circular equilibrium
states\cite{ring}, according to which the propagation speed $c_{_{\rm
E}}$ of extrinsic (wiggle) perturbations determines the effective
rotation velocity $v$, namely that of the current in the timelike
case, for which one obtains $v^2=T/U ={\cal L}/\Lambda$, and that of
the orthonormal tangent direction if the current is spacelike, in
which case one obtains $v^2=T/U=\Lambda/{\cal L}$.  Finally in the
``chiral'' case for which the current is null, both formulae are valid
simultaneously: one will have $\Lambda={\cal L}$ and $v=1$.

In all of these cases the vorton circumference will be given by
\begin{equation}
{\ell}_{\rm v}={|B|\over\sqrt{\cal - KL}} \ ,
\end{equation}
and the equilibrium condition to be solved for the state function of
the vorton will be expressible in the more directly utilisable form
\begin{equation}
{\cal K}^2{{\cal L}\over\Lambda} =b^2
\label{vorton}\end{equation}
using the abbreviation $b$ for the {\it Bernoulli ratio}, as defined
by
\begin{equation}
b={B\over C}=2\pi \kappa_{_0} \,{N\over Z}\ ,
\label{ratio}\end{equation}
where $N$ and $Z$ are the corresponding integer valued winding number
and particle quantum number (of which, if the current were
characterised by a non zero electric coupling constant $e$, the latter
would determine the vorton's total ionic charge, namely $Q=Ze$, as in
ordinary atomic physics).

From the well known theorem~\cite{stabgen} that (although there may be
instabilities with respect to non axisymmetric perturbations in
certain cases) the circular equilibrium states are {\it always} stable
with respect to perturbations that preserve their circular symmetry,
it follows that within the admissible range (\ref{stabc}) the
effective energy function ${\mit \Upsilon}$ can be extreme only at a
minimum but never at a maximum. This evidently implies that, within a
continuously connected segment of the admissible range, there can be
at most a single such extremum: in other words for a given value of
the conserved ratio $b^2$, the vorton equation can have {\it at most}
one solution for the state variable $w$ -- and hence for any function
thereof, such as the derived variable ${\cal K}$ and the corresponding
vorton circumference ${\ell}$, which will thus be {\it uniquely}
determined. It will be seen in the next section that in some cases
there will be no solution at all, i.e. there are values of the ratio
$b^2$ for which ${\mit \Upsilon}$ is monotonic throughout the allowed
range, so that a corresponding vorton state does not even exist.

\section{Solution of the equations of motion.}

The results in the immediately preceding section are independent of
the particular form of the Lagrangian ${\cal L}$. If we now restrict
ourselves to the specific case of the model (\ref{Lag}), we can use
the results of the earlier sections to rewrite the effective barrier
energy function ${\mit \Upsilon}$ in the form
\begin{equation}
{\mit \Upsilon}= m_\ast^{\,2}{\ell}\left(
\alpha+\ln\sqrt{\cal K}+\Big({C\over m_\ast{\ell}}\Big)^2 {\cal
K}\right)
\ .
\label{firstupsilon}
\end{equation} with ${\cal K}$ given explicitly as a function of
${\ell}$
by (\ref{ksolution}). A convenient way of applying this formula is to
think of ${\cal K}$ as the independent variable, with the
circumference ${\ell}$ (and hence the radius $r={\ell}/2\pi$) given by
\begin{equation}
m_\ast^{\, 2}{\ell}^2={{B}^2\!-{C}^2{\cal K}^2\over {\cal K}-1 }\ .
\label {lengthdep}\end{equation}
In the case $b^2<1$, which means ${B}^2<{C}^2$, this determines
${\ell}$ as a monotonically increasing function of ${\cal K}$ in the
{\it timelike} current range, $\hbox{e}^{-2\alpha}<{\cal K}<1$. In the
case $b^2>1$, which means ${B}^2>{C}^2$, this determines ${\ell}$ as a
monotonically decreasing function of ${\cal K}$ in the {\it spacelike}
current range, $1<{\cal K}<2$. In either case, we finally obtain the
effective barrier energy function in the form
\begin{equation}
{\mit \Upsilon}=m_\ast|{C}|\sqrt{b^2-{\cal K}^2\over
{\cal K}-1 }\left(\alpha+\ln\sqrt{\cal K}+{ {\cal K}({\cal K}-1)
\over b^2-{\cal K}^2 } \right)
\label{upsilon}\end{equation}
as a fully explicit function just of ${\cal K}$.  The formula
(\ref{deri}) for the derivative of this function gives
\begin{equation}
{d{\mit \Upsilon}\over d{\ell}}=m_\ast^2\left( \alpha+\ln\sqrt{\cal K}
-{b^2({\cal K}-1)\over {\cal K}(b^2-{\cal K}^2)}\right) \ .
\label{derivative}\end{equation}
It can thus be seen that the vorton equilibrium requirement
(\ref{vorton}) -- expressing the condition (\ref{equilibrium}) that
this derivative should vanish -- will be given for this particular
string model by
\begin{equation}
{\cal K}={{\cal K}_{\rm v}}
\end{equation}
where ${{\cal K}_{\rm v}}$ is obtained by solving the equation
\begin{equation}
b^2={{{\cal K}_{\rm v}}^{\, 2}(\alpha +\ln\sqrt{{{\cal K}_{\rm v}}})
\over
 \alpha-1+\ln\sqrt{{{\cal K}_{\rm v}}}+{{{\cal K}_{\rm v}}}^{-1} } \ .
\label{kvor}\end{equation}
Whenever an admissible solution exits, it can be seen that the
corresponding value
\begin{equation}
M={M_{\rm v}}
\end{equation}
of the mass of the vorton state will be given by
\begin{equation}
{{M_{\rm v}}\over m_\ast}=|{C}|\sqrt{{{\cal K}_{\rm v}}-1\over
b^2-{{\cal K}_{\rm v}}^{\, 2}}
\left({b^2\over{{\cal K}_{\rm v}}} +{{\cal K}_{\rm v}}\right) \ .
\label{mvor}\end{equation}

\section {The confinement effect and Classification of solutions}

Since ${\cal K}$ tends to unity for large values of $\ell$, it can be
seen from Eq.~(\ref{firstupsilon}) that the effective potential $\Upsilon$
grows linearly with radius at large distances. This means that no
matter
how large its energy may be, the loop can never expand to infinity: it
is subject to a confinement effect (not unlike that which motivated
early
attempts to use string models to account for the phenomenon of quark
confinement in hadron theory~\cite{Nambu}).

The fact that it admits no possibility of unbound trajectories
distinguishes the loop problem considered here from cases such as the
familiar example a point particle, of mass $m$ say, moving in the
Newtonian gravitational field of a central mass, $M_\ast$, say. 
In that case, the
orbits can be classified as Type 0, Type 1, and Type 2, where Type 0
means the special case of constant radius (circular) orbits, Type 1
means the generic case of varying radius but nevertheless bound
orbits, and Type 2 denotes unbound orbits. These types can be
subclassified into categories A and B, where A stands for ``always
regular'' or ``avoiding trouble'' and B stands for ``badly
terminating''. For Type 0 orbits, the ``good'' subcategory A is
clearly the only possibility. However while Type 1 orbits are
generically of type 1A, which for an inverse square law means the
elliptic case, there is also the possibility of type 1B orbits,
meaning bound trajectories of purely radially moving type, which end
by plunging into the central singularity. Similarly Type 2 orbits are
generically of type 2A, which for an inverse square law means the
parabolic and hyperbolic cases, but there is also the possibility of
type 2B orbits, meaning unbound trajectories of purely radially moving
type which begin or end at the central singularity. In the simple
point particle case the only relevant parameters are the orbital
binding energy $E$ say and the angular momentum $J$ say. Subcategory B
corresponds to the special case $J=0$. In the inverse square law case
the classification is simplified by the property of self symmetry with
respect to the transformations $E\mapsto E/s$, $J\mapsto J\sqrt s$
where $s$ is a scale factor: thus for the generic subcategory A, the
classification depends just on the invariant dimensionless combination
$EJ^2/m^3 M_\ast^{\, 2}$, being Type 0 for its absolute minimum value,
which is $-1/2$, Type 1 for a higher but still negative value, and
Type 2 otherwise.

The same principles can be applied to the classification of solutions
of the circular string loop problem, for which one only needs the Type
1 -- with ``good'' and ``bad'' subcategories 1A and 1B -- and the Type
0 -- which in this case means a vorton state, which can only be
``good''. There is no analogue of Type 2 for the string loop problem
because the possibility of an unbound orbit does not exist. This is
because the relevant effective potential function ${\mit \Upsilon}$
does not only diverge to infinity (due to the centrifugal effect) as
the radius $r$ becomes small, i.e. as ${\ell}\rightarrow 0$, which
corresponds to ${\cal K}\rightarrow |b|$: it is evident that ${\mit
\Upsilon}$ must also diverge (due to the energy needed for stretching
the string) in the large $r$ limit, i.e. as ${\ell}\rightarrow\infty$
which corresponds to ${\cal K}\rightarrow 1$.

Despite the fact that instead of the five possibilities (namely 0, 1A,
1B, 2A, 2B) needed or the point particle problem there are only three
(namely 0, 1A, 1B) in the circular string loop problem, the state of
affairs for this latter problem is considerably more complicated
because the orbits are not fully characterised just by the mass energy
parameter $M$ and the angular momentum parameter $J$: they also depend
on the Bernoulli constants ${B}$ and ${C}$ [which, by
(\ref{proportion}), are respectively proportional to the microscopic
winding number $N$ and the particle number $Z$]. According to
(\ref{angquant}) these constants are related by the condition
${B}{C}=2\pi J$, but that still leaves three independent parameters
which may conveniently be taken to be $M$, ${B}$, ${C}$ say -- instead
of the two that were sufficient for the point particle case. As in the
inverse square law case for a point particle, the flat space string
loop problem is self similar with respect to scale transformations,
which are expressible in this case by ${B}\mapsto {B} s$, and
${C}\mapsto {C} s$ and $M\mapsto M s$ (so that $J\mapsto Js^2$). Thus
whereas all that mattered qualitatively in the inverse square law was
a {\it single} dimensionless ratio (namely that between $J^2$ and
$E^{-1}$), in a corresponding manner the not so simple behaviour of
the circular string loop is qualitatively dependent on the {\it two}
independent dimensionless ratios relating ${B}^2$, ${C}^2$ and
$M^2$. A further complication is that the nature of this dependence
depends on the dimensionless parameter $\alpha$ characterising the
underlying string model.

Unlike the mass-energy parameter $M$, whose conservation depends on
the stationary character of the space time background, and would no
longer hold exactly when allowance is made for losses from
gravitational radiation, the winding number and particle number
satisfy conservation laws of a less conditional nature, so (although
their local conservation is also symmetry dependent) the corresponding
Bernoulli parameters ${B}$ and ${C}$ provide more fundamental
information about the string loop. It is therefore appropriate to use
their ratio $b$ as the primary variable in a classification of the
solution (with the understanding that $b=\infty$ means ${C}=0$).

Proceeding on this basis, the relevant parameter space can be
described in terms of five consecutive zones for the parameter
$b^2$. The reason why there are so many possibilities is that the
range of ${\ell}$, from $0$ to $\infty$ corresponds, according to
(\ref{lengthdep}), to a range of ${\cal K}$ from $1$ to $|b|$, which
may extend beyond the range (\ref{admissible}) that is physically
admissible according to the criterion (\ref{stabc}).

Between the limits where it diverges, ${\mit
\Upsilon}\rightarrow+\infty$, as ${\cal K}\rightarrow 1$ and ${\cal
K}\rightarrow |b|$, the effective potential energy function ${\mit
\Upsilon}$ will vary smoothly with at least one local minimum. However
according to the theorem recalled at the end of the previous section,
${\mit \Upsilon}$ can have at most one local minimum and no local
maximum within the admissible range (\ref{admissible}).  Moreover,
since $\alpha$ is strictly positive by its construction
(\ref{massrat}), it is evident that the large radius limit
$K\rightarrow 1$ will always lie safely within the physically
admissible range (\ref{admissible}). This leaves only two alternative
possibilities, which are either that ${\mit \Upsilon}$ should be
monotonic, with $d{\mit \Upsilon}/d{\ell}>0$, throughout the
physically admissible range (\ref{admissible}), or else that this
admissible range should include a turning point at a critical value of
${\ell}$ within which the derivative $d{\mit \Upsilon}/d{\ell}$ will
become negative, in which case it will have to remain negative all the
way to the inner limit of the admissible range. It is directly
apparent from the expression (\ref{derivative}) for $d{\mit
\Upsilon}/d{\ell}$ that there is no possibility for it to remain
positive near the limit of the admissible range in the timelike
current case, i.e. as $\ln\sqrt{\cal K}\rightarrow -\alpha$, so for
$b^2 <1$ the vorton equilibrium equation (\ref{kvor}) will always have
a physically admissible solution. However in the spacelike current
case, $b^2>1$, for which the relevant limit of the admissible range is
given (independently of $\alpha$) by ${\cal K}\rightarrow 2$, it can
be seen that it is indeed possible for the gradient (\ref{derivative})
to remain positive, the condition for this being the criterion for the
first of the qualitatively different zones listed as follows.

\smallskip
{\bf Zone I.\,} This is the ``fatal'' spacelike zone characterised by
\begin{equation}
b^2\left(1-{1\over 2\alpha +\ln 2}\right) \geq 4
\label{range1}\end{equation}
for which no admissible vorton solution exists. For such a scenario it
can be seen from (\ref{upsilon}) that the mass energy must necessarily
satisfy the condition
\begin{equation}
M\geq{M_{\rm s}}
\label{massbound}\end{equation}
where the mass limit ${M_{\rm s}}$ is given by
\begin{equation}
{{M_{\rm s}}\over m_\ast}= \sqrt{{B}^2-4{C}^2}\left(\alpha+{1\over
2}\ln 2
+{2\over b^2-4} \right) \ .
\label{msat}\end{equation}
In this case, after a possible phase of expansion to a maximum radius
obtained by solving ${\mit \Upsilon}=M$, the loop will inevitably
contract until it reaches the current saturation limit at ${\cal K}=2$
at which stage our classical string description will break down. This
means that in terms of the terminology introduced above, all Zone I
trajectories are of Type 1B. On Fig.~1 is displayed the potential
${\mit \Upsilon}$ against (top) the value of ${\cal K}$ and (bottom)
that of the circumference ${\ell}$, all quantities being rescaled with
the {\sl Kibble mass} $m$. It should be noted that for $\alpha\agt 1$,
the potential is roughly (i.e., up to negligible logarithmic
corrections) independent of $\alpha$ when seen as function of $\ell$
but not as a function of ${\cal K}$. This shows that the most relevant
parameter for cosmological applications is $m$ and not $m_\ast$ even
though the latter is essential for the very existence of vorton
states. The starting point of this zone marks the end of the curves on
Fig.~6.

It is to be remarked that in order for this zone to be of finite
extent, the carrier mass scale $m_\ast$ must not be too large compared
with the Kibble mass scale $m$, the precise condition being that the
value of $\alpha$ given by (\ref{massrat}) should satisfy the
inequality
\begin{equation}
\alpha>(1- \ln 2 )/2
\label{witcon}\end{equation}
If this condition were not satisfied -- which would be unlikely in a
realistic model, since the Witten mechanism cannot be expected to work
if the carrier mass is too large~\cite{witten,neutral,enon0} -- then
Zone I would consist only of the extreme limit $b^2=\infty$ i.e. the
case $C=0$, for which the string falls radially inwards with a
spacelike current but zero angular momentum.

\smallskip
{\bf Zone II.\,} This is the ``dangerous'' spacelike zone
characterised by
\begin{equation}
b^2 \geq 4> b^2\left(1-{1\over 2\alpha +\ln 2}\right)
\label{range2}\end{equation}
[which would consist of the entire range $b^2\geq 4$ if (\ref{witcon})
were not satisfied] for which the trajectory may be of (stationary)
Type 0, (well behaved oscillatory) Type 1A, or (badly behaved) Type
1B, depending on its energy. The Type 1B case is that for which $M$
satisfies an inequality of the form (\ref{massbound}), in which case
the loop will evolve in the same way as in the previous scenario, and
thus will again end up by contracting to a state of current
saturation. The ``good'' type 1A possibility, is characterised by the
condition that the mass should lie in the range
\begin{equation}
{M_{\rm s}}>M>{M_{\rm v}}
\end{equation}
where the maximum beyond which the current will ultimately saturate is
given by the preceding formula (\ref{msat}) for ${M_{\rm s}}$, and the
minimum value ${M_{\rm v}}$ is the mass of the relevant vorton state as
characterised by (\ref{mvor}): when this latter condition is satisfied
the loop will oscillate in a well behaved manner between a minimum and
a maximum radius that are obtained by solving ${\mit
\Upsilon}=M$. Finally the Type 0 possibility is that of the vorton
state itself, as given by the minimum value $M={M_{\rm v}}$. Similarly
to Fig.~1, Fig.~2 shows the potential in this zone II, against either
${\cal K}$ (top) and $\ell m/|C|$ (bottom), with the same remark as
before when $\alpha\agt 1$.

\smallskip
{\bf Zone III.\,} This is the ``safe'' zone characterised by
\begin{equation}
4>b^2>\hbox{e}^{-4\alpha}
\end{equation}
for which there is no danger of bad behaviour, i.e. the only
possibilities are the well behaved Type 1A, which applies to the
entire range
\begin{equation}
M>{M_{\rm v}}
\label{entire}\end{equation}
and the vortonic Type 0, as given by $M={M_{\rm v}}$.

It is to be remarked that this ``safe'' zone consists of three
qualitatively distinct parts, namely a subrange of spacelike current
solutions, Zone III\{+\} say (Fig.~3), given by
\begin{equation}
2>|b|>1\ ,
\end{equation}
a subrange of timelike current solutions, Zone III\{-\} say (Fig.~4),
given by
\begin{equation}
1>|b|>\hbox{e}^{-2\alpha}\ ,
\end{equation}
and in between the special ``chiral'' case of null current solutions,
Zone III\{0\} say, which is given just by $|b|=1$.

\smallskip
{\bf Zone IV.\,} This is the ``dangerous'' timelike zone characterised
by
\begin{equation}
\hbox{e}^{-4\alpha} \geq b^2>0
\label{IVdang}\end{equation}
for which (as in Zone II) the trajectory may be of (stationary ) Type
0, (well behaved oscillatory) Type 1A , or (badly behaved) Type 1B,
depending on its energy (see Fig.~5). The latter will occur whenever
\begin{equation}
M\geq{M_{\rm r}}\ ,
\label{lax}\end{equation}
where the relevant minimum mass -- above which the loop will contract
to a state of complete relaxation, i.e. zero tension -- is given by
\begin{equation}
{{M_{\rm r}}\over m_\ast}=|{C}|\hbox{e}^{-2\alpha}\sqrt{
1-\hbox{e}^{-2\alpha}\over  \hbox{e}^{-4\alpha} - b^2}  \ .
\end{equation}
The ``good'' type 1A possibility, is characterised by the condition
that the
mass should lie in the range
\begin{equation}
{M_{\rm r}}> M >{M_{\rm v}}
\end{equation}
where as before ${M_{\rm v}}$ is the vorton mass value given by
(\ref{mvor}), while finally the Type 0 possibility occurs when
$M={M_{\rm v}}$. It is to be remarked that the zone (\ref{IVdang})
includes a subzone characterized by the strict condition
$c_{_E}<c_{_L}$, where $c_{_L}^2=-dT/dU$ is the woggle velocity, a
condition that is expressible as~\cite{analytic} ${\cal K} <
\hbox{e}^{2(1-2\alpha )}$ and which has been conjectured to be
sufficient to ensure classical stability of the corresponding vorton
state.

\smallskip
{\bf Zone V.\,} This is the ``fatal'' timelike zone characterised by
\begin{equation}
b^2=0
\end{equation}
i.e. the case for which ${B}=0$ (to which the investigation of Larsen
and Axenides~\cite{LA} was devoted) for which it can be seen that the
mass must satisfy the inequality
\begin{equation}
{M\over m_\ast}>|{C}|\sqrt{1-\hbox{e}^{-2\alpha}} \ .
\end{equation}
(in which the lower bound is the common limit to which ${M_{\rm v}}$
and ${M_{\rm r}}$ converge as ${B}$ tends to zero). In this case (as
in the more extensive range covered by Zone I) the trajectory must be
of Type 1B, its ultimate fate being to reach a state of relaxation,
$T\rightarrow 0$, as in Zone IV when (\ref{lax}) is satisfied.

All these zones are shown on Fig.~6 where, as functions of the
parameter $|b|$ are plotted the value of the function ${\cal K}$ that
minimizes the potential in all but Zone I (top), the corresponding
value of the vorton mass $M_{_V}$ (middle) and length $\ell$ (bottom),
all in units of the Kibble mass $m$. It should be clear on this figure
that in most cases, the latter two are almost independent of $\alpha$,
the largest dependence occurring in zones II and IV.

\section{Conclusions}

In view of the potential cosmological interest of vorton formation, it
is of interest to distinguish the range of conditions under which a
cosmic string loop can survive in an ``A type'' oscillatory state --
that will ultimately damp down towards a stationary vorton
configuration -- from the alternative range of conditions under which
the loop will undergo a ``B type'' evolution, whereby it reaches a
configuration for which the classical string description breaks down,
in which case the investigation of its subsequent fate -- and in
particular of the question of whether the underlying vacuum vortex
defect will ultimately survive at all -- will need more sophisticated
methods of analysis than are presently available.

The present investigation is restricted to the case of exactly
circular loops for which it is shown, on the basis of the best
available classical string model\cite{analytic} that there is an
extensive range of parameter space, including the whole of Zone III in
the above classification, for which the ``A type'' solutions (that are
propitious for ultimate vorton formation) will indeed be obtained. On
the other hand it is also shown that (unlike what occurs in the
classical point particle problem) badly behaved ``B type'' solutions
are not limited to the special zero angular momentum case, Zone V, to
which a preceding study~\cite{LA} of this problem was restricted, but
are of generic occurrence, occupying the whole of Zone I and extensive
parts of Zones II and IV. It remains an open question whether these
results are representative of what will happen in the more general
case of initially non-circular loops.

The foregoing results are based on an analysis that is purely
classical in the sense that it neglects both quantum effects and also
the General Relativistic effects of the gravitational field. In
realistic cases of cosmological interest -- involving cosmic strings
produced at or below the GUT transition level -- it is to be expected
that the neglect of gravitational effects will be a very good
approximation: as remarked in the introduction, the relevant
Schwarzschild radius will usually be so small that the question of
black hole formation will be utterly academic, while the effect of
gravitational radiation, although it may become cumulatively
important, will be allowable for in the short run in terms of a very
slow ``secular'' variation of the mass parameter $M$, whereby an
oscillatory (Type 1A) trajectory will gradually settle down towards a
stationary (Type 0) vorton state.

Unlike the usually small corrections that will arise from gravitation,
the effects of quantum limitations may be of dominant importance for
realistic cosmological applications.  The preceding analysis should be
valid for loops characterised by sufficiently large values of the
winding number $N$ and particle quantum number $Z$, and thus for
correspondingly large values of the Bernoulli constants ${B}$ and
${C}$ and hence of $M$.  However it can be expected to break down
whenever the loop length ${\ell}$ becomes small enough to be
comparable with the Compton wavelength
\begin{equation}
{\ell}_\ast=m_\ast^{-1}
\end{equation}
associated with the carrier mass scale $m_\ast$.  It can be seen from
(\ref{lengthdep}) that the current saturation limit ${\cal
K}\rightarrow 2$ cannot be attained without violating the classicality
condition
\begin{equation}
{\ell}\gg{\ell}_\ast
\label{classical}\end{equation}
unless the corresponding dimensionless Bernoulli constants ${B}$ and
${C}$ [which by (\ref{proportion}) will have the same order of
magnitude provided $\kappa_{_0}$ is of order unity] are such as to
satisfy the condition
\begin{equation}
{B}^2-4{C}^2\agt 1   \ .
\end{equation}
This differs from the corresponding purely classical condition
${B}^2>4{C}^2$ (characterising Zones I and II) by having $1$ instead
of zero on the right hand side. It can similarly be seen that the
relaxation ($T\rightarrow 0$) limit, ${\cal K}\rightarrow
\hbox{e}^{-2\alpha}$, cannot be obtained without violating the
classicality
condition (\ref{classical}) unless
the Bernoulli constants satisfies the condition
\begin{equation}
{C}^2 \hbox{e}^{-4\alpha} - {B}^2   \agt
1-\hbox{e}^{-2\alpha} \ .
\end{equation}
which is similarly stronger than the corresponding purely classical
condition ${C}^2\hbox{e}^{-4\alpha} > B^2$ (characterising Zones IV
and V).

\section*{Acknowledgments}

It is a pleasure to thank D.W.~Sciama and E.P.S.~Shellard for many
stimulating discussions. P.P. would also like to thank the
International School for Advanced Studies at Trieste (Italy) for
hospitality during part of the time this work was being done.
A.G. thanks the Schuman foundation for the award of a fellowship, and
acknowledges partial financial support from the programme
Antorchas/British Council (project 13422/1--0004).

\figure{Fig.~1: The potential ${\mit \Upsilon} /(m |C|)$ as a function
of
${\cal K}$ (top) and $m {\ell} /|C|$ (bottom) for zone I. Here and on
the following figures, it is found that for $\alpha \agt 1$, the
curves as functions of $m \ell /|C|$ all coincide (up to negligible
logarithmic corrections) so they can be shown for different values of
$\alpha$ ranging from 1 to 100 by the same thick curve. (Note that
this simplification depends on normalising with respect to the Higgs
mass $m$ rather than the carrier mass scale $m_\ast$).  It is clear
however that the variations with ${\cal K}$ are strongly dependent on
the ratio $\alpha$.}

\figure{Fig.~2: Same as Fig.~1 for zone II.}
\figure{Fig.~3: Same as Fig.~1 for zone III\{ -\}.}
\figure{Fig.~4: Same as Fig.~1 for zone III\{ +\}.}
\figure{Fig.~5: Same as Fig.~1 for zone IV. In this zone, for large
values of $\alpha$, the minimum value of the potential is attained
only for very small values of ${\cal K}$ and hence are not visible on
the figure.}

\figure{Fig.~6: The vorton state function ${\cal K}_{_V}$, mass
$M_{_V}/ m|C|$ and length $m \ell /|C|$ against the Bernoulli ratio
$|b|$. From $\alpha = (1-\ln 2)/2$ to $\alpha =1$, the curve is
smoothly deformed from the long-dashed one to the thick one which
includes many values of $\alpha$, showing explicitly the independence
in $\alpha$.}

\onecolumn 

\begin{figure}[t]
\begin{center}
\leavevmode
\epsfxsize=6in \epsfbox{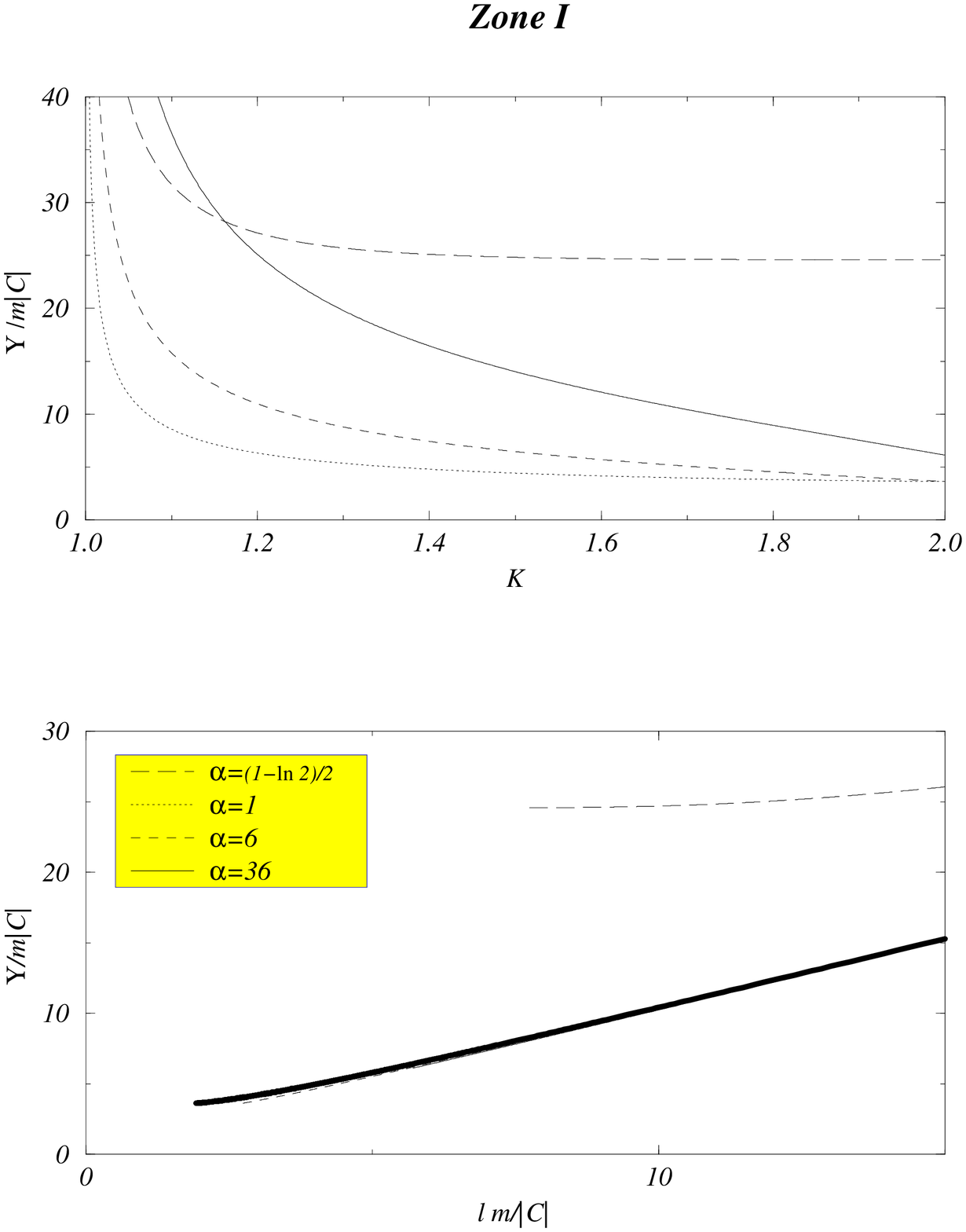}
\end{center}
\end{figure}

\newpage

\begin{figure}[t]
\begin{center}
\leavevmode
\epsfxsize=6in \epsfbox{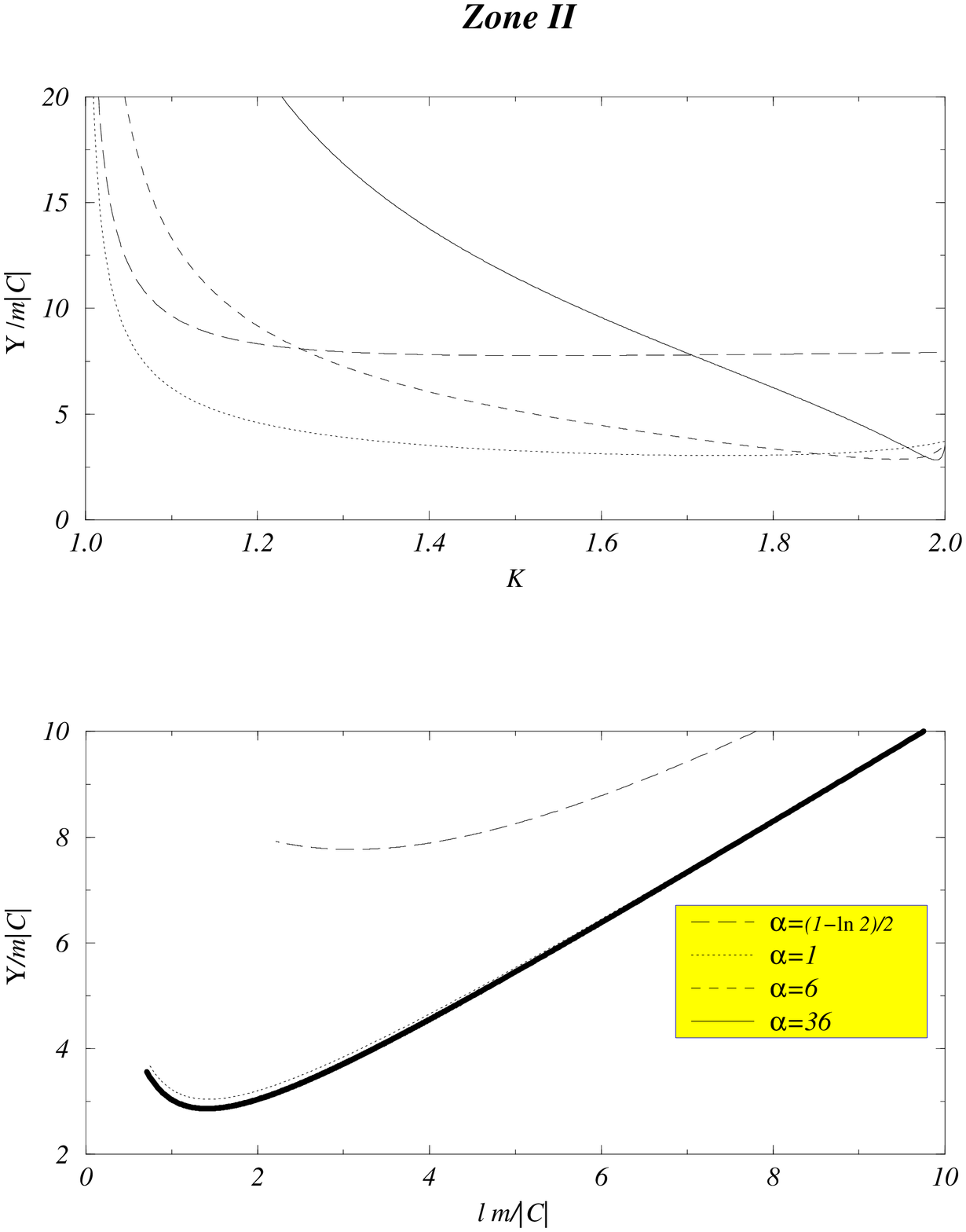}
\end{center}
\end{figure}

\newpage

\begin{figure}[t]
\begin{center}
\leavevmode
\epsfxsize=6in \epsfbox{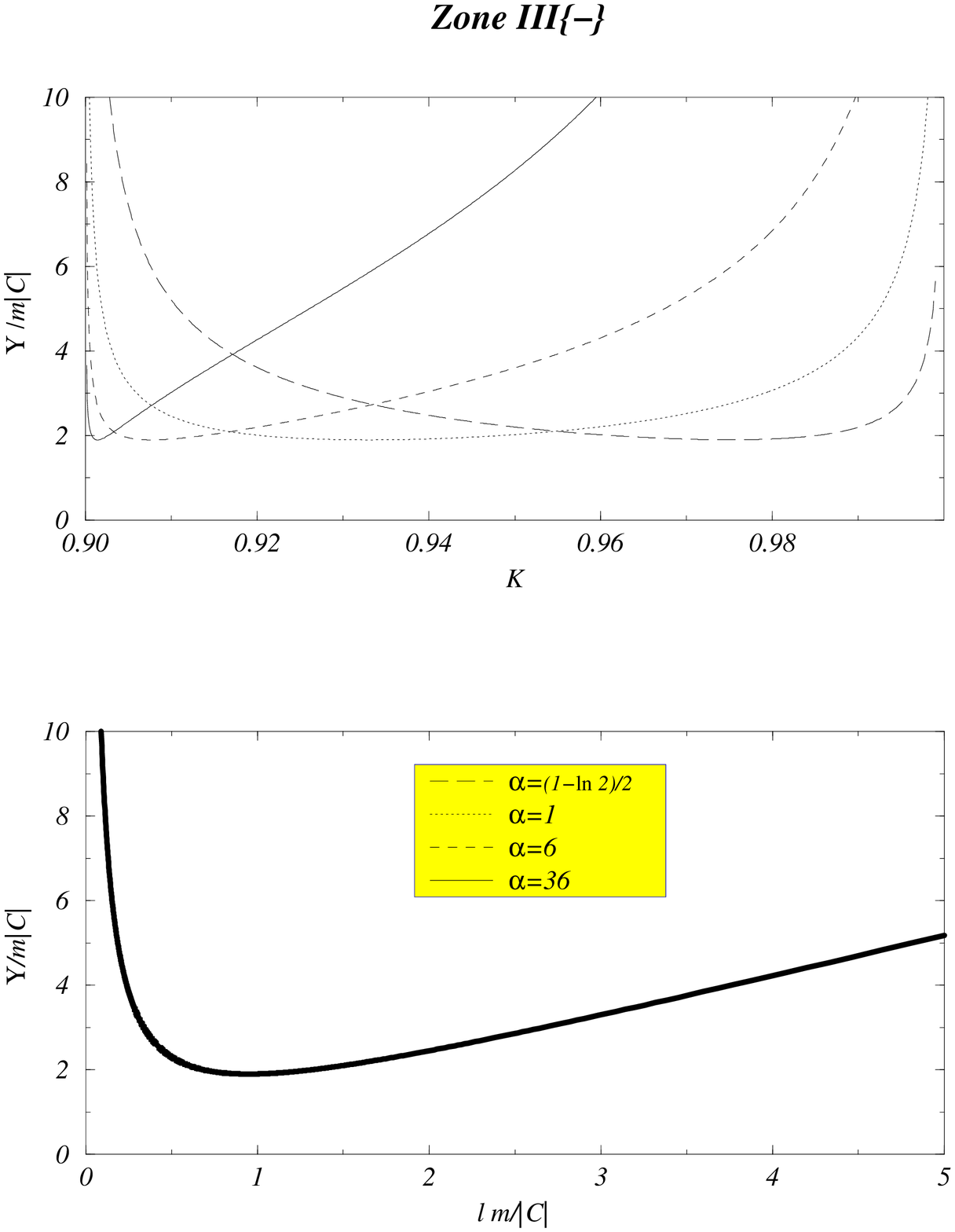}
\end{center}
\end{figure}

\newpage

\begin{figure}[t]
\begin{center}
\leavevmode
\epsfxsize=6in \epsfbox{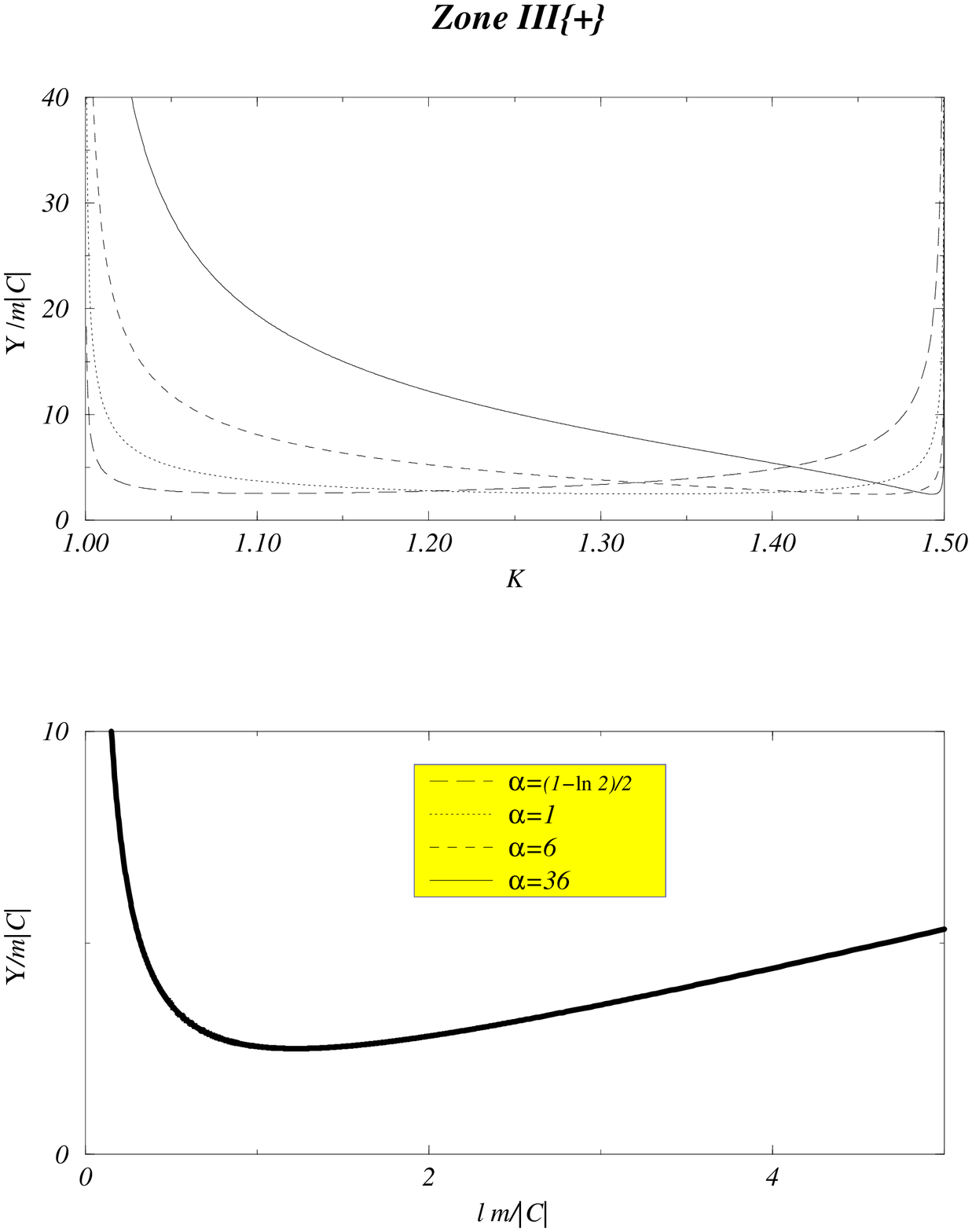}
\end{center}
\end{figure}

\newpage

\begin{figure}[t]
\begin{center}
\leavevmode
\epsfxsize=6in \epsfbox{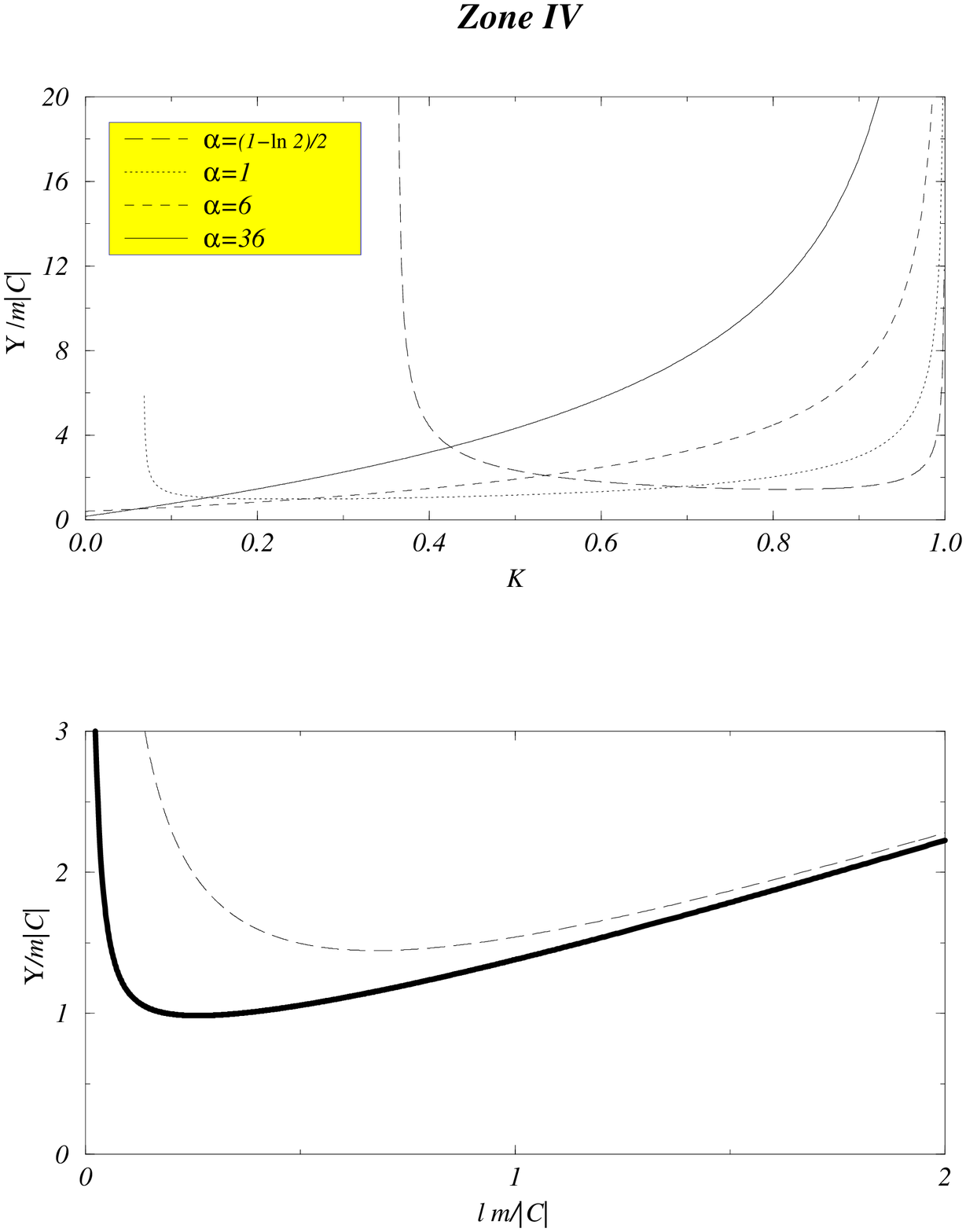}
\end{center}
\end{figure}

\newpage

\begin{figure}[t]
\begin{center}
\leavevmode
\epsfxsize=6in \epsfbox{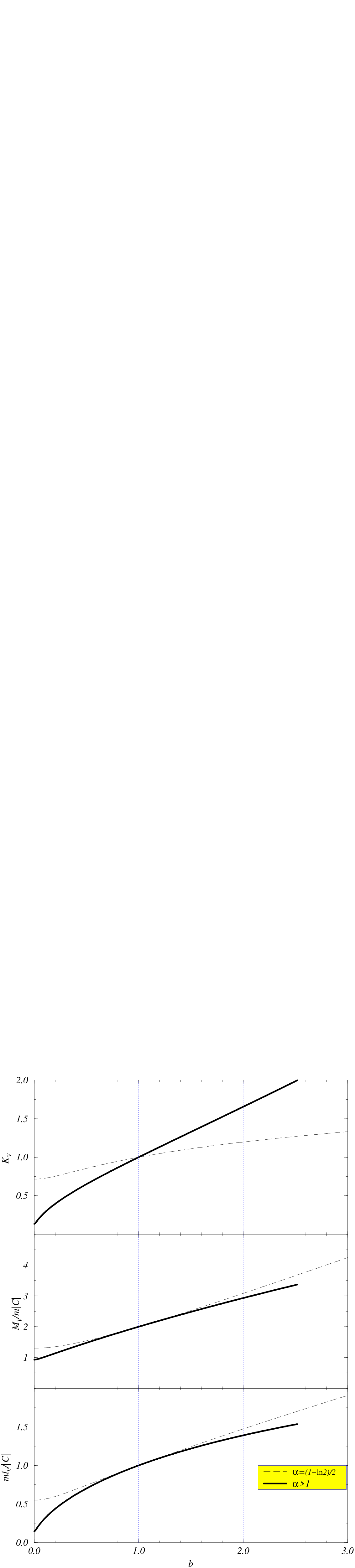}
\end{center}
\end{figure}


\begin{references}

\bibitem{kibble} T.~W.~B.~Kibble, J. Math. Phys. A {\bf 9}, 1387
(1976), Phys. Rep. {\bf 67}, 183 (1980).

\bibitem{book} E.~P.~S.~Shellard \& A.~Vilenkin, {\it Cosmic strings
and other topological defects}, Cambridge University Press (1994).

\bibitem{DS} R.~L.~Davis, E.~P.~S.~Shellard, \prd {\bf 38}, 4722
(1988); Nucl. Phys. B {\bf 323}, 209 (1989);

\bibitem{witten} E.~Witten, Nucl. Phys. B {\bf 249}, 557 (1985).

\bibitem{vorton} B.~Carter, Ann. N.Y. Acad.Sci., {\bf 647}, 758 (1991);
B.~Carter, Proceedings of the XXXth Rencontres de Moriond,
Villard-sur-Ollon, Switzerland, 1995, Edited by B.~Guiderdoni and
J.~Tran Thanh V\^an (Editions Fronti\`eres, Gif-sur-Yvette, 1995).

\bibitem{96014} R.~Brandenberger, B.~Carter, A.-C.~Davis, M.~Trodden,
{\it Cosmic vortons and particle physics constraints}, preprint No
96014 (Observatoire de Paris), hep-ph/9605382, DAMTP-96-16,
BROWN-HET-1036, MIT-CTP-2515

\bibitem{for} B.~Carter, Phys. Lett. {\bf B 224}, 61 (1989)
and {\bf B 228}, 466 (1989).

\bibitem{mal} B.~Carter, in {\it Formation and Interactions of
Topological Defects} (NATO ASI {\bf B349}), ed R. Brandenberger \&
A-C. Davis, {\it pp} 303-348 (Plenum, New York, 1995).

\bibitem{neutral} P.~Peter, \prd {\bf 45}, 1091 (1992).

\bibitem{enon0} P.~Peter, \prd {\bf 46}, 3335 (1992).

\bibitem{stabgen} B.~Carter, X.~Martin, Ann. Phys. {\bf 227}, 151
(1993).

\bibitem{stab} X.~Martin, \prd {\bf 50}, 749 (1994).

\bibitem{stabwit} X.~Martin, P.~Peter, \prd {\bf 51}, 4092 (1995).

\bibitem{SWH} S.~W.~Hawking, Phys. Lett. {\bf B 231}, 237 (1989).

\bibitem{LA} A.~L.~Larsen, M.~Axenides, {\it Runaway collapse of
Witten vortex loops}, preprint, Alberta Thy 13-96, hep-th/9604135.

\bibitem{BGP} B.~Boisseau, H.~Giacomini and D.~Polarski, \prd {\bf 51},
6909 (1995).

\bibitem{nielsen} N.K.~Nielsen, Nucl. Phys.  {\bf B167}, 249 (1980);
N.K. Nielsen, P.Olesen, Nucl. Phys. {\bf B291}, 829 (1987).

\bibitem{several} A.~Davidson and K.~Wali, Physics Letters, {\bf B213},
439 (1988); A.~Davidson, N.K.~Nielsen, and Y.~Verbin, Nucl. Phys. {\bf
B412}, 391 (1994).

\bibitem{larsen} A.L.~Larsen, Class. Quantum. Grav. {\bf 10}, 1541
(1993).

\bibitem {CX} B. Carter, \prl {\bf 74}, 3093 (1995);
X. Martin, \prl {\bf 74}, 3102 (1995).

\bibitem{analytic} B.~Carter, P.~Peter, \prd {\bf 52}, R1744 (1995).

\bibitem{wall} P.~Peter, {\it Current-carrying domain walls}, J. Phys.
{\bf A} (1996) in press.

\bibitem{alford} M.~Alford, K.~Benson, S.~Coleman, J.~March-Russell,
Nucl. Phys. {\bf B 349}, 414 (1991).

\bibitem{EW} P.~Peter, \prd {\bf 46}, 3322 (1992)

\bibitem{warm} B.~Carter, Nucl. Phys. {\bf B412}, 345 (1994)

\bibitem{CFH} B.~Carter, V.P.~Frolov, O.~Heinrich, Class. Quantum.
Grav.,
{\bf 8}, 135 (1991).

\bibitem{brane} B.~Carter, Class. Quantum. Grav. {\bf 9S}, 19 (1992).

\bibitem{ring} B.~Carter,  Phys. Lett. {\bf 238}, 166 (1990).

\bibitem{Nambu} Y.~Nambu, \prd {\bf 10}, 4262 (1974).

\end{references}
\end{document}